\documentclass[11pt]{article}
\usepackage{amsmath,amssymb,amsfonts}
\usepackage[nosort]{cite}
\usepackage{graphicx,color}
\usepackage[colorlinks=true, linkcolor=blue, citecolor=blue, linktoc=all]{hyperref}
\usepackage[usenames,dvipsnames]{xcolor}
\usepackage{sfmath}
\usepackage{tikz-cd}

\usepackage[makeroom]{cancel}
\usepackage{slashed}
\newcommand{\solder}{\theta}
\newcommand{\dsolder}{P}
\newcommand{\mX}{\mathfrak{X}}
\newcommand{\mY}{\mathfrak{Y}}

\newcommand{\mD}{\mathfrak{D}}
\newcommand{\mg}{\mathfrak{g}}
\newcommand{\ghost}{c}

\newcommand{\hatd}{\hat{d}}
\newcommand{\un}[1]{\underline{#1}}

\newcommand{\cM}{M}

\newcommand{\chkM}{{\color{red} \,\checkmark\kern-5pt{}_{M}}}
\newcommand{\cX}{\mathfrak{X}}
\newcommand{\be}{\begin{equation}}
\newcommand{\ee}{\end{equation}}
\newcommand{\beq}{\begin{eqnarray}}
\newcommand{\eeq}{\end{eqnarray}}
\newcommand{\bea}{\begin{eqnarray}}
\newcommand{\eea}{\end{eqnarray}}
\newcommand{\beqn}{\begin{eqnarray}}
\newcommand{\eeqn}{\end{eqnarray}}
\newcommand{\secE}{\underline{\psi}}
\newcommand{\basE}{\underline{e}}
\def\pa{\partial}

\newcommand{\hlt}[1]{{\color{WildStrawberry}{\em #1}}\index{#1}}

\newcommand{\thistitle}{Lie Algebroids and the Geometry of Off-shell BRST}
\newcommand{\ulb}[1]{
	\centerline{
		\begin{minipage}[c]{0.7\textwidth}
			\begin{center}
			${}^{#1}$ Physique Math\'ematique des Interactions Fondamentales \& International Solvay Institutes, Universit\'e Libre de Bruxelles, Campus Plaine - CP 231, 1050 Bruxelles, Belgium
			\end{center}
		\end{minipage}
		}
	}
\newcommand{\uiuc}[1]{
	\centerline{
		\begin{minipage}[c]{0.7\textwidth}
			\begin{center}
			${}^{#1}$ Illinois Center for Advanced Studies of the Universe \& Department of Physics,\\ 
			University of Illinois, 1110 West Green St., Urbana IL 61801, U.S.A.
			\end{center}
		\end{minipage}
		}
	}

\begin{document}

%\allowdisplaybreaks
\title{\thistitle}
\author{
	Luca Ciambelli$^{a}$ and Robert G. Leigh$^{b}$
	\\
	\\
	{\small \emph{\ulb{a}}} \\ \\
	{\small \emph{\uiuc{b}}}
	\\
	}
\date{\today}
\maketitle
\vspace{-5ex}
\begin{abstract}
\vspace{0.4cm}
It is well-known that principal bundles and associated bundles underlie the geometric structure of classical gauge field theories. In this paper, we explore the reformulation of gauge theories in terms of Lie algebroids and their associated bundles. This turns out to be a simple but elegant change, mathematically involving a quotient that removes spurious structure. The payoff is that the entire geometric structure involves only vector bundles over space-time, and we emphasize that familiar concepts such as BRST are built into the geometry, rather than appearing as adjunct structure. Thus the formulation of gauge theories in terms of Lie algebroids provides a fully off-shell account of the BRST complex. We expect that this formulation will have appealing impacts on the geometric understanding of quantization and anomalies, as well as entanglement in gauge theories.  The formalism covers all gauge theories, and we discuss Yang-Mills theories with matter as well as gravitational theories explicitly.
\end{abstract}

\newpage

\begingroup
\hypersetup{linkcolor=black}
\tableofcontents
\endgroup
\noindent\rule{\textwidth}{0.6pt}

\setcounter{footnote}{0}
\renewcommand{\thefootnote}{\arabic{footnote}}
\newpage

\section{Introduction}

The geometrization of gauge theories in terms of principal fibre bundles, originally initiated in \cite{10.1143/PTP.16.537, Lubkin:1963zz, PhysRevD.12.3845, RevModPhys.52.175}, is well-established and gives a precise mathematical structure that has many applications in both physics and mathematics. 
In this interpretation, a principal connection on a principal bundle  corresponds to a gauge field and charged matter fields  to sections of associated bundles. 
One of the central ideas in gauge theories is BRST symmetry \cite{Becchi:1975nq, Tyutin:1975qk}, which underlies quantization and both the physical spectrum of the theories as well as the structure of quantum anomalies \cite{Gribov:1977wm, Bonora:1982ve,  Gregoire:1992tf, Grigoriev:2006tt, Upadhyay:2010ww}.\footnote{The literature on this subject is vast. A list of reviews, focussed on different aspects of BRST, may be found in \cite{Nemeschansky:1987xb, Henneaux:1989jq,Henneaux:1992ig, Becchi:1996yh, VanHolten:2001nj, Fuster:2005eg}.} 
There is a deep mathematical structure underlying each of these involving the cohomology of the nilpotent operator generating the BRST symmetry, deeply related to Lie algebra cohomology \cite{Chevalley:1948zz}; a necessarily non-exhaustive list of references concerning BRST cohomology includes \cite{Zumino:1984ws, Bandelloni:1986wz, Henneaux:1989rq, Brandt:1989gv, Henneaux:1995ex, Dragon:1996md, DuboisViolette:1991is, Barnich:1994db, Barnich:1994mt, Barnich:1995ap, Brandt:1996mh, Barnich:2000zw}.

 In physics discussions, the BRST structure is described by employing Grassmann-valued ``ghost'' fields. However, it has long been appreciated that there is an alternative, more geometric, viewpoint in which the ghost fields are nothing but partners of the gauge fields that arise naturally in the principal bundle framework.
This idea was originally formulated by Thierry-Mieg and collaborators, \cite{Neeman:1979cvl, ThierryMieg:1979xe, ThierryMieg:1979kh, Baulieu:1981sb, ThierryMieg:1982un, ThierryMieg:1987um}. Subsequently, this idea was extended and improved upon \cite{Bonora:1980pt, Bonora:1980ar, Bonora:1981rw, Bonora:1981rv, Bonora:1982ef,Hirshfeld:1980te, Quiros:1980ec, Delbourgo:1981cm, Hoyos:1981pb, Hull:1990qg}.
 The main point here is that (the reader should think here of Yang-Mills theories, but we will present the general case including gravity in the body of the paper) a connection on a principal fibre bundle $P\to M$ can be thought of as a section of $T^*P$ valued in the Lie algebra, whereas in physics language, a gauge field is thought of as a section of $T^*M$ (a one-form) valued in the Lie algebra. Thus the gauge field is only the ``horizontal'' part of the connection, the ``vertical'' part proposed to correspond to the ghost field. This works out because the Grassmann algebra is isomorphic to an exterior differential algebra, and so the ghosts can be alternatively represented as differential forms on the principal bundle.

Although we regard this as a beautiful suggestion, nevertheless the Grassmann language is almost always used in place of this geometric construction. Presumably, this is because the mathematics involved is somewhat formidable, and the Grassmann formulation simpler to understand. One of the things that we would like to emphasize in this paper (and indeed repair) is that the principal bundle construction obscures some physics because of a technical point: many of the interesting features of gauge theories involve 
in this formalism the tangent and cotangent bundles $TP$ and $T^*P$, which are bundles over $P$ itself rather than space-time $M$. This presents complications for understanding local physics on $M$, and in addition in some sense there is superfluous structure present. This comes about because the gauge group acts both on the left and on the right. The left action corresponds to the familiar gauge transformations, while the right action is redundant. While this redundancy is well-understood, it complicates the presentation of the theory. The subject that we want to  explore in this paper recenters the discussion of gauge theories in terms 
 of the principal bundle (and its tangent and cotangent bundles) to the corresponding \hlt{Atiyah Lie algebroid}, first developed in \cite{Atiyah:1957, Atiyah:1979iu}. The Atiyah Lie algebroid is precisely the quotient of $TP$ by the right action, and so understanding it has the benefit of the redundancies being  removed.  Lie algebroids in general have many rather technical applications in mathematical physics \cite{MR0214103, MR0216409, zbMATH01186367, kosmannschwarzbach2002differential, fern2007lie, Ramandi:2014qoa,marle2008differential}, but are perhaps unfamiliar to most readers. In the interest of making the paper self-contained, we will carefully review the subject of Lie algebroids. This is by no means the first time that Lie algebroids have been advocated for formulating gauge theories
\cite{Lazzarini_2012, Fournel:2012uv, Jordan:2014uza, Carow-Watamura:2016lob, Kotov:2016lpx, Attard:2019pvw} (or for deforming them \cite{Strobl:2004im, Bojowald:2004wu, Mayer:2009wf}), but we do not believe that the significance of BRST in this framework has been appreciated. 
It is part of our agenda in this paper to make the material, while mathematically rigorous,  more accessible to the physics reader, while correcting and extending some of the previous work. 
The primary mathematical foundations of the subject are contained in the two books \cite{mackenzie_1987,mackenzie_2005} by Kirill Mackenzie.\footnote{We were saddened to hear of the recent passing of Professor Mackenzie. The present work would not have been possible but for his important contributions to this mathematical field.}

We have several motivations for advocating Lie algebroids in this context. First, they resolve the aforementioned technical problems: the Lie algebroid is a vector bundle over space-time (along with some extra structure that we will specify later), and thus does not contain, by construction, the superfluous degrees of freedom. In a very precise sense, the fibres of the (Atiyah) Lie algebroid correspond to vectors in space-time and generators of the Lie algebra of the structure group of $P$. 
Thus the Lie algebroid efficiently encodes both the diffeomorphisms of the base space-time $M$ as well as gauge transformations, and leads to a fully covariant formalism. The significance of the fact that the Atiyah Lie algebroid is a vector bundle over $M$ (as is any associated bundle)  rather than $P$ should not be underestimated. It makes possible the understanding of all physically relevant concepts in terms of maps between vector bundles, which then have a direct local interpretation on the base manifold $M$. This includes both connections on the Lie algebroids
as well as matter fields, represented as sections of associated bundles. As we will establish, connections on a Lie algebroid acquire a simple and effective geometrical meaning as Ehresmann connections. Here, no extra requirement of right invariance is imposed, for the latter concept is already taken care of by definition of Atiyah Lie algebroids. 

We will show how BRST transformations are encoded in the structure of the Lie algebroid: they are  part of the exterior derivative, which we call $\hatd$, on the Lie algebroid, going hand in hand with the usual exterior derivative on $M$. This structure aligns well with the above-stated usual BRST formalism, in which one considers extended forms, formal sums of differential forms on $M$ and Grassmann-valued ghosts.  
Indeed, it is well-known that the corresponding extended exterior derivative combining ordinary derivatives and BRST transformations is nilpotent and thus has a cohomology of physical significance. In the formalism that we will review and develop in this paper, the corresponding concept, $\hatd$, is shown to include the standard BRST transformations and as such the latter are geometrically well-defined. We will show that the Fadeev-Popov gauge ghost fields \cite{Faddeev:1967fc} arise as part of the connections on the Lie algebroid and all of the familiar properties of the aforementioned extended formalism, such as the ``Russian formula'', are automatic. In the formalism involving principal bundles directly, such properties are obtained upon restricting attention to $G$-invariant quantities, whereas in Grassmann implementations of BRST, it is a required input.
%rather than being assumptions. 
It should be emphasized that this BRST structure, since it is encoded in the geometrical structure, is ``off-shell'', rather than being associated with any particular (classical) action or gauge-fixing mechanism. Given a particular geometric structure, we can instead ask what (classical) actions are consistent with it, being well-defined top forms on $M$ that can be integrated. Such possibilities are in fact encoded in the cohomological structure of $\hatd$. We do not address the introduction of particular dynamics in this paper, but one should expect additional structure, such as antighosts, to be necessary. These fields are not required in our off-shell formalism. 
Therefore, the key result of our work is the construction of a framework where classical gauge theories and their BRST structure are completely geometrized,  without having extra unphysical degrees of freedom to remove with external restrictions on the geometry.

The paper is organized as follows. In Section \ref{sec:LieAlg}, we review the construction of transitive Lie algebroids. This section includes discussions of the maps involved in defining the Lie algebroid and their relation to the connection. We emphasize the (global) split of the Lie algebroid into horizontal and vertical sub-bundles. As an example, we review the construction of derivations on vector bundles as having the structure of a Lie algebroid. This construction serves to define a representation of a given Lie algebroid on associated vector bundles, leading to the definition of the exterior derivative $\hatd$ on the Lie algebroid and associated bundles. We emphasize in this discussion the horizontal-vertical split of $\hatd$ and we have chosen notation such that this split corresponds to the usual physics notation of a covariant derivative of a section of an associated bundle along a direction in $M$ (the horizontal part) and the BRST transformation (the vertical part). We complete this section with a discussion of this novel BRST interpretation.

In Section \ref{sec:atiyah}, we provide an account of the standard mathematical formulation of gauge theories, employing principal connections on principal gauge bundles. We emphasize that this formulation suffers from the presence of the redundant right action of the structure group, and that taking the quotient of $TP$ by this right action leads to an Atiyah Lie algebroid over $M$. As an example of a transitive Lie algebroid, all of the results of Section \ref{sec:LieAlg} carry over immediately. We finish this section with a discussion of the significance of the construction to physics, briefly reviewing the aforementioned Thierry-Mieg idea and starting a discussion on physical data that will culminate at the end of Section \ref{Sec4}, once local fields are carefully introduced.

Sections \ref{sec:LieAlg} and \ref{sec:atiyah} were presented in a coordinate- and basis-independent language by making use of arbitrary auxiliary sections of the various vector bundles.  The benefit of this is that the formalism is automatically well-defined, and fully diffeomorphism- and gauge-invariant. In Section \ref{Sec4}, we describe the local structure of Atiyah Lie algebroids, to make contact with the usual physics ``index'' notation. We fully describe how the connection on an Atiyah Lie algebroid encodes locally a gauge field as well as a gauge ghost field and show how, by carefully constructing the local trivialization, the usual gauge transformations and diffeomorphisms are recovered. We end Section \ref{Sec4} with a discussion of the physics of classical gauge theories and how the Lie algebroid structures neatly encode them. We should acknowledge at this point that essentially all of the geometric structure described here rests on the standard notion of connections on vector bundles (and we have made a significant effort to clarify that fact). However, with the extra geometric structure associated with the Lie algebroid, the physical interpretation is apparently novel, and naturally includes off-shell BRST.

In Section \ref{sec:grav}, we show how the Atiyah Lie algebroid structure can be used to describe standard gravitational theories in the first order formalism. The basic geometric object is an Atiyah Lie algebroid constructed from the frame bundle (or more precisely, a related $G$-structure).\footnote{Original accounts of $G$-structures are \cite{chern1966, molino1972, Godina_2003}, see also the standard book reference \cite{kobayashi1995transformation}.} Along with the connection, we describe how the usual notion of a solder form can be carried over naturally to a section of a bundle associated to this Atiyah Lie algebroid. From this point of view, the gravitational theory is essentially equivalent to any other Yang-Mills theory, albeit with a non-compact structure group. Finally in Section \ref{sec:diffghosts}, we discuss the absence of diffeomorphism ghosts in the Lie algebroid formalism and how this fits with the more traditional BRST constructions. 
 
In Section \ref{sec:concl}, we conclude with remarks and discussion of planned followup works. Appendix \ref{sec:notation} is devoted to an account of the conventions used throughout the paper, while Appendix \ref{app:LAmaps} discusses the horizontal-vertical split of $\hatd$ acting on general forms valued in associated vector bundles. 

\section{Lie Algebroids}\label{sec:LieAlg}

In this section, we introduce the mathematical structure behind the notion of transitive Lie algebroids and begin the study of its consequences in physics. We start by defining the abstract notion of transitive Lie algebroids and how a horizontal distribution is constructed via an Ehresmann connection. Then, we introduce a specific such algebroid: the bundle of derivations of an associated bundle $E$. This in turn allows us to discuss the cochain complex and the action of its exterior derivative on forms valued in $E$. Singling out the horizontal and complementary vertical pieces of the exterior derivative is crucial for physical application, as we carefully establish at the end of this section, where we show that the BRST transformation of physical fields is built into this mathematical construction. The conventions adopted in this section are summarized in Appendix \ref{sec:notation}, to which we direct the reader  unfamiliar with the general differential geometry abstract notation that we employ. The theory of transitive Lie algebroids is extensively studied in Refs. \cite{mackenzie_1987, mackenzie_2005}, which are useful references for this section, together with \cite{Lazzarini_2012, Fournel:2012uv}.

\subsection{Transitive Lie algebroids}\label{sec:TLA}

%A transitive Lie algebroid is a vector bundle $A\to M$ over a manifold $M$ along with a (surjective) \hlt{anchor map} $\rho:A\to TM$. 
A Lie algebroid is a vector bundle $A\to M$ over a manifold $M$ along with an  \hlt{anchor map} $\rho:A\to TM$. In the following, we will confine our attention to transitive Lie algebroids, where the anchor map is surjective. This allows us to introduce short exact sequences, crucial for our construction. It may well be interesting 
to explore
intransitive Lie algebroids as well, although we do not do so here.
For now we will not specify $A$, but we will consider specific cases relevant to physics later in the paper. We will refer to sections of $A$ as $\un\mX,\un\mY,...$; these are not themselves vector fields, but the anchor map serves to project them onto ordinary vector fields, as $\rho(\un\mX)\in\Gamma(TM)$. Note that we will use the sections $\un\mX,...$ and vector fields $\un X,...$ as auxiliary objects; they will not be interpreted as physical fields. In $TM$, we have the usual Lie bracket ${\cal L}_{\un X}\un Y=\left[\un X,\un Y\right]$, and we suppose that we also have a Lie bracket
 on $A$, denoted $\left[\un\mX,\un\mY\right]_A$. The Lie bracket on $A$ satisfies the Leibniz rule via the anchor map. That is, for $f,g\in C^\infty(M)$,
\beq\label{rhoLeib}
\left[f\un\mX,g\un\mY\right]_A=f g\left[\un\mX,\un\mY\right]_A+f\rho(\un\mX)(g)\  \un\mY-g\rho(\un\mY)(f)\  \un\mX,
\eeq
where, since $\rho(\un\mX)\in\Gamma(TM)$, $\rho(\un\mX)(g)\in C^\infty(M)$ is the ordinary derivative of $g$ along $\rho(\un\mX)$.
%It then follows from the Jacobi identity that the anchor map gives a representation of $TM$ in $A$
It then follows from the Jacobi identity that the anchor map gives a representation of $A$ in $TM$; equivalently $\rho$ is a morphism,\footnote{Here, and throughout the paper, when we refer to a map being a morphism, we mean that it preserves the Lie bracket, as in eq. \eqref{rhorep}. As we will see, this is a crucial concept in this subject. The quantity $R^\rho$ will be referred to as the curvature of the map $\rho$, and we will employ similar notation for other bundle maps that will arise in the following.} satisfying
\beq\label{rhorep}
R^\rho(\un\mX,\un\mY)=\left[\rho(\un\mX),\rho(\un\mY)\right]-\rho(\left[\un\mX,\un\mY\right]_A)=\un 0.
\eeq
Therefore the Lie bracket on $A$ is not linear (with respect to multiplication of $\un\mX$ or $\un\mY$ by functions), but only in the sense that the usual Lie bracket on $TM$ is non-linear, involving differentiation of the functions $f,g$. 

The image of the anchor map $\rho$ is $TM$, and in general, $\rho$ will have a kernel, 
\[ \ker\rho=\left\{ \un\mX\in\Gamma(A) \Big| \rho(\un\mX)= \un 0\right\},\]
which is itself a vector bundle over $M$. Often $\ker\rho$ is referred to as the \hlt{vertical sub-bundle}  $V\subset A$.
Formally, we introduce a vector bundle $L$ over $M$, and an injective map $\iota:L\to A$, whose image is $\ker\rho$, and thus $\rho\circ\iota=0$ (i.e., $\rho(\iota(\un\mu))=\un 0$, $\forall \un\mu\in\Gamma(L)$). This implies $\ker\rho=im(\iota)=V$. Thus we have all the elements of a short exact sequence
\beq\label{shortExactSeqGenLieAlgd}
\begin{tikzcd}
0
\arrow{r} 
& 
L
\arrow{r}{\iota} 
& 
A
\arrow{r}{\rho} 
& 
TM
\arrow{r} 
&
0
\end{tikzcd}.
\eeq
Given that this is exact, the tangent bundle is a quotient, $TM=A/V$. 
We will denote sections of $L$ as $\un\mu,\un\nu$ and the Lie bracket on $L$ as $\left[\cdot,\cdot\right]_L$. 
We require the map $\iota$ to be a morphism,
\beq
R^\iota(\un\mu,\un\nu)=\left[\iota(\un\mu),\iota(\un\nu)\right]_A-\iota(\left[\un\mu,\un\nu\right]_L)=\un 0,\qquad\forall \un\mu,\un\nu\in\Gamma(L).
\eeq
Note that \eqref{rhoLeib} then implies that the bracket on $L$ is linear,
\beq\label{iotalinbrac}
\left[f\un\mu,g\un\nu\right]_L=fg\left[\un\mu,\un\nu\right]_L
\eeq and so the Lie bracket on $L$ defines a Lie algebra (above each point in $M$).
So we see that $A$ can be regarded 
as a vector bundle whose fibres at each point contain ordinary vectors tangent to $M$, as well as vectors in a Lie algebra $L$. There is an interesting integrability problem in which a Lie algebroid is ``exponentiated'' to a Lie groupoid, but as far as we understand, this remains an active area of mathematical research \cite{fern2007lie}. In the present paper, we will not make use of Lie groupoids, but we expect that they will be of interest in physics for certain problems at least.

\subsubsection{Connections on Lie algebroids and the horizontal bundle}\label{sec:LAconns}

We have seen above that given a Lie algebroid $A$, the tangent bundle may be thought of as the quotient $TM=A/V$.
Vector fields in $TM$ are thus in one-to-one correspondence with equivalence classes of sections of $A$, with the equivalence being $\un\mX\sim\un\mX+\un\mY_V$, with $\un\mY_V\in \Gamma(V)$.  There is no unique choice of a representative of each class; making such a choice is equivalent to introducing a connection on a Lie algebroid, in the sense of Ehresmann. This defines a split of the short exact sequence, specifying how to lift vector fields in $TM$ to sections of $A$. As such, it defines the \hlt{horizontal sub-bundle} $H\subset A$ such that $A=H\oplus V$ globally, where $V$ is the aforementioned vertical sub-bundle. We will call  $\sigma: TM\to H$ the map realizing this lift. As such, we have
\beq\label{rhosigid}
\rho\circ\sigma:\un X\mapsto\rho\circ\sigma(\un X)=\rho(\sigma(\un X))=\un X,\qquad\forall \un X\in\Gamma(TM).
\eeq
That is, \beq\label{rhosigid}\rho\circ\sigma=Id_{TM},\eeq whereas $\sigma\circ\rho:A\to A$ acts as a projector\footnote{Here, we are are implying that $\sigma\circ\rho$ acts as an isomorphism on $H$. Consequently, since $(\sigma\circ\rho)^2=\sigma\circ\rho$, we have  $\sigma\circ\rho\big|_H=Id_H$.} on $A$, \[\sigma\circ\rho(\un\mX)\equiv\un\mX_H\in \Gamma(H),\quad H\subset A.\] Generally, the connection $\sigma$ will have curvature, which we can express as
\beq
\un 0\neq R^\sigma(\un X,\un Y)=\left[\sigma(\un X),\sigma(\un Y)\right]_A-\sigma(\left[\un X,\un Y\right])\in\Gamma(A).
\eeq
So for  Lie algebroids, one way to express the concept of curvature of a connection is simply the failure of $\sigma$ to be a morphism. We will soon introduce other notions of curvature for a Lie algebroid, but as we will show,  they will all turn out to be equivalent.
 
Notice that $R^\sigma$ is vertical, that is $R^\sigma(\un X,\un Y)\in \Gamma(V), \forall \un X,\un Y\in\Gamma(TM)$.   Indeed, if we map it to $TM$ using $\rho$, we find
\beqn\label{Rsigisvert}
\rho(R^\sigma(\un X,\un Y))=\rho(\left[\sigma(\un X),\sigma(\un Y)\right]_A)-\rho\circ\sigma(\left[\un X,\un Y\right])=\left[\rho\circ\sigma(\un X),\rho\circ\sigma(\un Y)\right]-\left[\un X,\un Y\right]
=\un 0,
\eeqn
where we used \eqref{rhorep} and \eqref{rhosigid}, and so $R^\sigma(\un X,\un Y)$ is in the kernel of $\rho$. 

Another important ingredient in the theory of Lie algebroids is the \hlt{connection reform}\footnote{The term reform is sometimes used in the mathematics literature  \cite{mackenzie_1987,mackenzie_2005}, and we use it here to help distinguish it from the usual notion of differential forms, sections of $T^*M$. We will later see in detail how the connection reform is related to the usual notion of connection forms in gauge theory.} $\omega: A\to L$. 
Equivalently we can regard $\omega\in\Gamma(A^*\times L)$, where $A^*$ is the dual bundle to $A$, whose sections are ``extended forms'',\footnote{We refer to sections of $\wedge^nA^*$ as extended forms because 
%in a sense that we will make precise, 
they consist of both horizontal and vertical forms, the horizontal forms being in one-to-one correspondence with differential forms on $M$. %The term extended form 
This terminology is also used in the BRST formalism where the extension is to formally add differential forms and Grassmann quantities. Here, the vertical parts of our extended forms will play the role of the Grassmann quantities. %We will also later introduce an extended exterior derivative $\hatd$, and this  will split into a horizontal (covariant) derivative and the BRST operator.
} dual to sections of $A$. Regarded as a map from $A$ to $L$, we require the connection reform to have $\ker(\omega)=im(\sigma)=H\subset A$. Thus we can now draw
\beq\label{splitshortExactSeq}
\begin{tikzcd}
0
\arrow{r} 
& 
L
\arrow{r}{\iota} 
\arrow[bend left]{l} 
& 
A
\arrow{r}{\rho} 
\arrow[bend left]{l}{\omega}
& 
TM
\arrow{r} 
\arrow[bend left]{l}{\sigma}
&
0
\arrow[bend left]{l} 
\end{tikzcd}.
\eeq
The composition $\omega\circ\iota$ is (minus)\footnote{The minus sign appearing here is chosen so as to ensure consistency with standard notation, such as eq. \eqref{defcurvatureform} below.} the identity on $L$, whereas $\iota\circ\omega$ is the projector on $A$ whose image is $V$. Thus, we now have two projectors on $A$, one for $H$ and one for $V$. For any section $\un\mX$ of $A$, we can decompose it into horizontal and vertical parts,
\beq\label{HVsplit}
\un\mX=\sigma\circ\rho(\un\mX)-\iota\circ\omega(\un\mX)\equiv\un\mX_H+\un\mX_V,
\eeq
with
\beq\label{HVsplitsforvectors}
\un\mX_H\equiv\sigma\circ\rho(\un\mX),\qquad \un\mX_V\equiv -\iota\circ\omega(\un\mX).
\eeq
Note that these formulae are consistent: for example, applying $\omega$ to \eqref{HVsplit} gives $\omega(\un\mX)$ on the left and $\omega\circ\sigma\circ\rho(\un\mX)-\omega\circ\iota\circ\omega(\un\mX)$ on the right, which coincides with $\omega(\un\mX)$ since $\omega\circ\sigma=0$ and 
\beq\label{omioid}\omega\circ\iota=-Id_L.\eeq
These formulae are a basis-independent way to project onto the horizontal and vertical, and we will make extensive use of them throughout the paper. The split performed here is also discussed in \cite{Lazzarini_2012, Fournel:2012uv}, with a similar set of conventions for the various bundle maps. It is important to note the following properties of the Lie bracket on $A$ with respect to this decomposition
%. Namely, we have
\beq\label{HValgebraprops}
[\un\mX_H,\un\mY_V]_A\in \Gamma(V),\qquad 
[\un\mX_V,\un\mY_V]_A\in \Gamma(V),
\eeq
which follow directly from \eqref{rhorep} and $\rho\circ\iota=0$.
The Lie bracket of two horizontal sections, $[\un\mX_H,\un\mY_H]_A$ has no vertical part if and only if $H$ is an integrable distribution in $A$; we will see later that integrability is equivalent to the vanishing of the curvature of the connection.

The connection reform $\omega$ has a curvature which we will be able to define once we have introduced the (nilpotent) exterior derivative $\hat d$ on $A$. As we will explicitly see below, this exterior derivative is not the usual de Rham exterior derivative $d$ associated with $M$, but is its analogue for $A$, mapping $\hatd:\wedge^r A^*\to \wedge^{r+1}A^*$. 

\subsection{Derivations on vector bundles}\label{sec:derivations}

To work towards an understanding of $\hatd$, we introduce a first example of a Lie algebroid, the derivations of a vector bundle. 
Given any vector bundle $E\to M$, we can introduce a sequence of related bundles of maps denoted $Diff^n(E)$ as follows. By $Diff^0(E)$ is just meant the endomorphisms\footnote{\label{foot:notation} Given a basis of sections $\basE_a$ for $E$ and $\basE_a\otimes f^b$ for $End(E)$, we have 
\beq
\varphi(\secE)=\varphi^a{}_b \basE_a\otimes f^b(\psi^c\basE_c)=(\varphi^a{}_b\psi^b) \basE_a,\qquad \varphi\in End(E),\quad \secE\in\Gamma(E)
\eeq
so the endomorphisms act as linear transformations of the components of the section of $E$. Here, we have taken $f^b$ as the dual basis of sections of $E^*$ so that $f^b(\basE_c)=\delta^b{}_c$. For more details and conventions concerning bases,  see Sec. \ref{Sec4} and App. \ref{sec:notation}. Note that for brevity we do not distinguish between the bundle $End(E)$ and its space of sections $\Gamma(End(E))$. We also note that conventionally, we do not underline sections of the bundles $End(E)$, $Diff^1(E)$ and $Der(E)$.} of the bundle, $Diff^0(E)=End(E)$, which have the property of linearity;  that is,  for $f\in C^\infty(M)$,  $\secE\in\Gamma(E)$ and $\varphi\in End(E)$, 
\beq\label{endolin}
\varphi(f\secE)=f\varphi(\secE).
\eeq 
The bundle $End(E)$ has a Lie bracket given by the commutator of endomorphisms, which is skew and linear, $\left[\varphi,f\varphi'\right]=f\left[\varphi,\varphi'\right]$. In what follows, $End(E)$ will play the role of $L$.

A first order differential operator $\Gamma(Diff^1(E))\ni D:E\to E$ is such that 
\beq
D(f\secE)-fD(\secE)=\varphi_f(\secE),\qquad \varphi_f\in End(E).
\eeq 
We can interpret this as Leibniz if the endomorphism $\varphi_f$ involves a derivative of the function $f$ associated to $D$. To do so, one introduces an (anchor) map $\rho_E$ to $TM$, defining a sub-bundle $Der(E)$ of $Diff^1(E)$.
That is, we require $\rho_E(\mD)$, for $\mD\in \Gamma(Der(E))$, to be an ordinary derivative on functions, $\varphi_f=\rho_E(\mD)(f)\in C^\infty(M)\subset End(E)$. The bundle $Der(E)$ together with the anchor map $\rho_E$ has the structure of a Lie algebroid.
So we define a \hlt{derivation} $\mD\in \Gamma(Der(E))$ such that 
\beq\label{derlinfunc}
\mD(f\secE)=f\mD(\secE)+\rho_E(\mD)(f)\ \secE.
\eeq 
The Lie bracket on $Der(E)$ is just given by composition 
\beq
\left[\mD,\mD'\right](\secE)=\mD(\mD'(\secE))-\mD'(\mD(\secE)).
\eeq
This is itself a derivation because
\beqn
\left[\mD,\mD'\right](f\secE)&=&\mD(\mD'(f\secE))-\mD'(\mD(f\secE))
\nonumber\\
&=&\mD(f\mD'(\secE))+\mD(\rho_E(\mD')(f)\ \secE)-\mD'(f\mD(\secE))-\mD'(\rho_E(\mD)(f)\ \secE)
\nonumber\\
&=&f\left[\mD,\mD'\right](\secE)+\left[\rho_E(\mD),\rho_E(\mD')\right](f)\ \secE
\nonumber\\
&=&f\left[\mD,\mD'\right](\secE)+\rho_E(\left[\mD,\mD'\right])(f)\ \secE,
\eeqn
where in the last line we used the fact that the anchor map is a morphism.
Finally, notice that sections of $\ker\rho_E$ are endomorphisms (in that case, \eqref{derlinfunc} becomes \eqref{endolin}), and so we see that indeed $Der(E)$ has the Lie algebroid structure
\beq
\begin{tikzcd}
0\arrow{r}
&
End(E)
\arrow{r}{\iota_E}
&
Der(E)
\arrow{r}{\rho_E}
&
TM
\arrow{r}
&
0\end{tikzcd}.
\eeq
As remarked upon above, we will make use of this construction extensively, once we have an idea of which vector bundles $E$ may be relevant to a given situation. That $Der(E)$ is a Lie algebroid has been explained in \cite{mackenzie_1987,mackenzie_2005}; see also e.g., \cite{Lazzarini_2012}. 
In fact, the bundle $Der(E)$ gives a representation of a Lie algebroid $A$ if we supply maps $\phi_E:A\to Der(E)$ and $v_E:L\to End(E)$.  In such a situation there will be a derivation associated to each section $\un\mX$ of $A$, as graphically shown hereafter
\beq\label{DerEshortExactSeqGen}
\begin{tikzcd}
0
\arrow{r} 
& 
L
\arrow{r}{\iota} 
\arrow{dd}{v_E}
& 
A
\arrow{dr}{\rho} 
\arrow{dd}{\phi_E} 
& 
&
\\
&&&
TM
\arrow{r} 
&
0
\\
0
\arrow{r} 
& 
End(E)
\arrow{r}{\iota_E} 
& 
Der(E)
\arrow{ur}{\rho_E} 
& 
&
\end{tikzcd}.
\eeq
We will later discuss our prime examples, Atiyah Lie algebroids, which are related to principal bundles, and in that case, the vector bundles $E$ will be associated bundles carrying a representation of the structure group. The map $v_E$ gives, for each section of $L$, an endomorphism, expressing the infinitesimal action of the group on sections of $E$ via the usual matrix representations. But before specializing, let us explore some important properties of such representations.

The central property of the maps $v_E,\phi_E$ is that they are morphisms.
That is,
\beq
[\phi_E(\un\mX),\phi_E(\un\mY)]=\phi_E([\un\mX,\un\mY]_A),\qquad
[v_E(\un\mu),v_E(\un\nu)]=v_E([\un\mu,\un\nu]_L).
\eeq
Using the notation for bases introduced in footnote \ref{foot:notation}, for an element $\un\mu=\mu^A\un t_A$, with $\{\un t_A\}$ a basis of sections of $L$ (we again refer to App. \ref{sec:notation} for further details), we write
\beq
v_E(\un\mu)=\mu^Av_E(\un t_A)=\mu^A (t_A)^a{}_b \basE_a\otimes f^b\equiv \mu^a{}_b \basE_a\otimes f^b.
\eeq
The fact that $v_E$ is a morphism simply translates into the fact that the $(t_A)^a{}_b$ give a matrix representation of the Lie algebra (see footnote \ref{foot:Liealg} below).

Similarly, since $\phi_E(\un\mX)\in\Gamma(Der(E))$ for any $\un\mX\in\Gamma(A)$, it acts as a derivation on sections of $E$. So we have
\beq\label{derlinfuncphi}
\phi_E(\un\mX)(f\secE)=f\phi_E(\un\mX)(\secE)+\rho(\un\mX)(f)\ \secE,\qquad \forall \secE\in\Gamma(E),\ f\in C^\infty(M),\ \un\mX\in\Gamma(A),
\eeq 
where $\rho_E\circ\phi_E=\rho$. 
Note that although we have described some of its features,  we have not actually fixed a definition of the  map $\phi_E$; we will supply such a definition later in the paper.  In the rest of this section, we will show, as we claimed above, that  $\phi_E$ must be a morphism of Lie brackets. 

Given that $\phi_E(\un\mX)$ is a derivation, it is natural to introduce a corresponding exterior derivative on $E$, defined as
\beq\label{defphiDE}
\phi_E(\un\mX)(\secE)\equiv (\hatd\secE)(\un\mX)
,\qquad \secE\in\Gamma(E),
\eeq
where we interpret $\hatd\secE: A\to E$, or equivalently, as a section $\hatd\secE\in\Gamma(A^*\times E)$ where $A^*$ is the bundle dual to $A$. We will now show that the fact that $\phi_E$ is a morphism translates into the nilpotency of $\hatd$. To extend the action of $\hat d$ to $E$-valued (extended) forms, we note that the latter are elements $\secE_n\in\Gamma(\wedge^n A^*\times E)$. Therefore, in analogy with the de Rham complex, we refer to $\Gamma(\wedge^n A^*\times E)$ as $\Omega^n(A,E)$ and introduce the cochain complex of (extended) forms in $A$ valued in $E$ as
\beq
\Omega^\bullet (A,E)=\bigoplus \Omega^n(A,E),
\eeq
where the sum goes from $0$ to the rank of $A$, and $\Omega^0(A,E)=\Gamma(E)$. The exterior derivative $\hat d$ is then the map $\hat d: \Omega^n(A,E)\to \Omega^{n+1}(A,E)$, and its action is explicitly given by the \hlt{Koszul formula}
\beqn\label{koszulE}
(\hatd\secE_n)(\un\mX_1,...,\un\mX_{n+1})
&\equiv&\sum_{r=1}^{n+1}(-1)^{r+1}\phi_E(\un\mX_r)(\secE_n(\un\mX_1,...,\widehat{\un\mX_r},...,\un\mX_{n+1}))
\\
&&+\sum_{r<s}(-1)^{r+s}\secE_n(\left[\un\mX_r,\un\mX_s\right]_A,\un\mX_1,...,\widehat{\un\mX_r},...,\widehat{\un\mX_s},..,\un\mX_{n+1}),
\nonumber
\eeqn
where overhats refer to omission.
The right-hand side is computable from \eqref{defphiDE} because $\secE_n(\un\mX_1,...,\un\mX_{n})\in\Gamma(E)$. It reduces to eq. \eqref{defphiDE} for $n=0$, and the second line of \eqref{koszulE} is such as to make the formula linear in all the sections $\un\mX_1,...,\un\mX_{n+1}$ (that is, only $\secE_n$ is differentiated).
We can immediately establish that $\hatd^2=0$ as follows: we simply write \eqref{koszulE} for $\secE_n=\hatd\secE_{n-1}$, and iterate. Let us do this explicitly for $n=1$:
\beqn\label{hatd2psi0}
(\hatd\hatd\secE_0)(\un\mX_1,\un\mX_{2})
&=&\phi_E(\un\mX_1)(\hatd\secE_0(\un\mX_2))-\phi_E(\un\mX_2)(\hatd\secE_0(\un\mX_1))
-\hatd\secE_0(\left[\un\mX_1,\un\mX_2\right]_A)
\nonumber\\
&=&\left[\phi_E(\un\mX_1),\phi_E(\un\mX_2)\right](\secE_0)
-\phi_E(\left[\un\mX_1,\un\mX_2\right]_A)(\secE_0).
\eeqn
{\it Thus we see that the nilpotency of $\hatd$ is equivalent to $\phi_E$ being a morphism.} 
Similarly, we can extend this computation to rank-$n$ $E$-valued forms, although the computations become tedious. Here we give just one more example, 
\beqn\label{hatd2psi0}
(\hatd\hatd\secE_1)(\un\mX_1,\un\mX_{2},\un\mX_{3})
&=&\phi_E(\un\mX_1)(\hatd\secE_1(\un\mX_2,\un\mX_3))
-\phi_E(\un\mX_2)(\hatd\secE(\un\mX_1,\un\mX_3))
+\phi_E(\un\mX_3)(\hatd\secE_1(\un\mX_1,\un\mX_2))
\nonumber
\\&&
-\hatd\secE_1(\left[\un\mX_1,\un\mX_2\right]_A,\un\mX_3)
+\hatd\secE_1(\left[\un\mX_1,\un\mX_3\right]_A,\un\mX_2)
-\hatd\secE_1(\left[\un\mX_2,\un\mX_3\right]_A,\un\mX_1)
\nonumber
\\
&=&\secE_1(\left[\left[\un\mX_1,\un\mX_2\right]_A,\un\mX_3\right]_A)
+\secE_1(\left[\left[\un\mX_3,\un\mX_1\right]_A,\un\mX_2\right]_A)
+\secE_1(\left[\left[\un\mX_2,\un\mX_3\right]_A,\un\mX_1\right]_A)
=\un 0,
\nonumber
\eeqn
which follows again from the facts that $\phi_E$ is a morphism and that the Lie bracket on $A$ satisfies the Jacobi identity.
This can be carried out to higher order, and so we conclude that $\hatd^2=0$ on the full cochain complex $\Omega^\bullet(A,E)$, if $\phi_E$ is a morphism.
\bigskip

\subsubsection{$\hatd$ for sections of $L$}\label{sec:hatdpsi}

Later, 
we will define $\phi_E$ for arbitrary associated vector bundles. But given a connection on $A$, we have an $L$-valued form, the connection reform $\omega$. The above technology can be applied by regarding $L$ as an example of a vector bundle $E$. Applying $\hatd$ to $\omega$, we get an $L$-valued two-form (i.e., an element of $\Omega^2(A,L)$); contracting with sections $\un\mX,\un\mY$ we have a section of $L$ which can be  mapped to $A$ using $\iota$. To do all of this explicitly, we need to understand $\phi_L$. In this case it is simply given by the Lie bracket on $A$ \cite{mackenzie_1987,mackenzie_2005}; given $\un\mu\in\Gamma(L)$ and $\un\mX\in\Gamma(A)$ we have\footnote{Eq. \eqref{Lderrep} is essentially the ad action; in \cite{mackenzie_1987,mackenzie_2005}, the $\iota$ map on both sides of the equation was implicit. Here we have included it for consistency. }
\beq\label{Lderrep}
\iota(\phi_L(\un\mX)(\un\mu))=[\un\mX,\iota(\un\mu)]_A.
\eeq
That this defines a morphism is equivalent to the Jacobi identity for the Lie bracket on $A$, that is
\beqn
\iota([\phi_L(\un\mX),\phi_L(\un\mY)](\un\mu))&=& 
\iota(\phi_L(\un\mX)(\phi_L(\un\mY)(\un\mu)))
-\iota(\phi_L(\un\mY)(\phi_L(\un\mX)(\un\mu)))
\\&=&
[\un\mX,[\un\mY,\iota(\un\mu)]_A]_A
-[\un\mY,[\un\mX,\iota(\un\mu)]_A]_A
\\
&=&
[[\un\mX,\un\mY]_A,\iota(\un\mu)]_A
\\
&=&
\iota(\phi_L([\un\mX,\un\mY]_A)(\un\mu)).
\eeqn
Applying $\hatd$ to $\omega$ gives
\beqn
\iota((\hatd\omega)(\un\mX,\un\mY))&=&
\iota\Big(\phi_L(\un\mX)(\omega(\un\mY))-\phi_L(\un\mY)(\omega(\un\mX))-\omega([\un\mX,\un\mY]_A)\Big)
\\
&=& [\un\mX,\iota\circ\omega(\un\mY)]_A-[\un\mY,\iota\circ\omega(\un\mX)]_A-\iota\circ\omega([\un\mX,\un\mY]_A),\label{firstresulthatdomega}
\eeqn
where we used \eqref{koszulE} for $n=1$ and \eqref{Lderrep} repeatedly. Making use of eqs. (\ref{HVsplitsforvectors}-\ref{HValgebraprops}) along with $R^\iota=0$, we then compute
\beqn\label{omgcurvature}
\iota((\hatd\omega)(\un\mX,\un\mY)
+[\omega(\un\mX),\omega(\un\mY)]_L)=
-\iota\circ\omega([\un\mX_H,\un\mY_H]_A).
\eeqn
Note that the right-hand side is the vertical part of the bracket $[\un\mX_H,\un\mY_H]_A$, which encodes the non-integrability of $H\subset A$. It is well-known (see e.g. \cite{mackenzie_1987, mackenzie_2005}) that this non-integrability is proportional to the curvature of the Ehresmann connection $\sigma$. Indeed, we have immediately using \eqref{HVsplit} and $R^\rho=0$
\beq
-\iota\circ\omega([\un\mX_H,\un\mY_H]_A)
=[\sigma\circ\rho(\un\mX),\sigma\circ\rho(\un\mY)]_A-\sigma\circ\rho([\un\mX_H,\un\mY_H]_A)
=R^\sigma(\rho(\un\mX),\rho(\un\mY)).
\eeq
An explicit realization of the Ehresmann connection (in the context of Atiyah Lie algebroids discussed in the upcoming sections) corresponds to the physics notion of a gauge field, and the usual notion of the curvature of the gauge field coincides with the curvature of the Ehresmann connection. 

Furthermore, we note that we can rewrite \eqref{omgcurvature} as
\beqn\label{omgcurvature2}
\Omega(\un\mX,\un\mY)=
-\omega([\un\mX_H,\un\mY_H]_A),
\eeqn
if we define
\beq\label{defcurvatureform}
\Omega=\hatd\omega+\frac12[\omega,\omega],\qquad \Omega\in\Gamma(\wedge^2A^*\times L)
\eeq
(given $\frac12[\omega,\omega](\un\mX,\un\mY)=[\omega(\un\mX),\omega(\un\mY)]_L$). We will refer to $\Omega$ as the \hlt{curvature reform} (that is, the curvature 2-form of the connection reform). 
In the context of Atiyah Lie algebroids discussed in the next section, we will see that the connection reform $\omega$ contains not only the physics notion of gauge field, but also encodes the notion of a BRST ghost field.

It is convenient to now introduce\footnote{We use the notation $R^\omega$ here because it vanishes if $-\omega$ is a morphism.}
\beq\label{defRomega}
R^\omega(\un\mX,\un\mY)\equiv [\omega(\un\mX),\omega(\un\mY)]_L+\omega([\un\mX,\un\mY]_A),
\eeq
in terms of which we may write \eqref{omgcurvature2} as
\beq\label{OmegaRomega}
\Omega(\un\mX,\un\mY)=-R^\omega(\un\mX_H,\un\mY_H).
\eeq
We will see below that $R^\omega$ also contains further information.
 
Consequently, there are multiple ways to express the curvature, 
\beqn\label{Russianformula}
R^\sigma(\rho(\un\mX),\rho(\un\mY))=\iota(\Omega(\un\mX,\un\mY))=-\iota(R^\omega(\un\mX_H,\un\mY_H)).
\eeqn
Given either expression, we see immediately that only the horizontal parts of the sections $\un\mX,\un\mY$ contribute; that is, $\Omega$ is a horizontal form in that $\Omega(\un\mX_V,\un\mY)=\un 0$ because $\rho\circ\iota=0$. For gauge theories, this result will translate to the \hlt{Russian formula} 
%which in the usual discussions of BRST geometry is introduced as a requirement 
(see for instance \cite{Neeman:1979cvl, ThierryMieg:1979xe, ThierryMieg:1979kh, Baulieu:1981sb, ThierryMieg:1982un, Baulieu:1984pf, Baulieu:1983tg, Baulieu:1985md, ThierryMieg:1987um}); here we see that it follows immediately from the conditions $R^\iota=0=R^\rho$, which are built-in properties of any Lie algebroid. Curvature of a Lie algebroid can be understood as the failure of $\sigma$ and $-\omega$ to be morphisms.
 
 \bigskip
 
To understand how we might choose $\phi_E$ for general bundles, let us explore further $\phi_L$.  
First we note that \eqref{Lderrep} implies
\beq\label{LbdlderivL}
\phi_L(\un\mX)(\un\mu)=-\omega([\un\mX,\iota(\un\mu)]_A),
\eeq
since $\omega\circ\iota=-Id_L$.  
It is useful to split $\un\mX$ appearing in \eqref{LbdlderivL} into horizontal and vertical parts,
\beqn
\phi_L(\un\mX)(\un\mu)
&=&-\omega([\un\mX_H,\iota(\un\mu)]_A)-\omega([\un\mX_V,\iota(\un\mu)]_A)
\\
&=&-\omega([\un\mX_H,\iota(\un\mu)]_A)
-[\omega(\un\mX_V),\un\mu]_L.\label{LbdlderivLsplit}
\eeqn
We can rewrite the second term in \eqref{LbdlderivLsplit} in terms\footnote{\label{foot:Liealg} For a bundle $E$, we have the morphism $v_E:L\to End(E)$, as in eq. \eqref{DerEshortExactSeqGen}. For $\lbrace\un t_A\rbrace$ a basis of sections of $L$, see App. \ref{sec:notation}, $v_E(\un t_A)$ is the corresponding matrix representation, and we have
\beqn
\left[v_E(\un t_A),v_E(\un t_B)\right] &=& ((t_A){}^c{}_{d} (t_B){}^d{}_{f}-(t_B){}^c{}_{d} (t_A){}^d{}_{f})\un e_c\otimes  f^f\\
v_E(\left[\un t_A,\un t_B\right]_L)&=&f_{AB}{}^Cv_E(\un t_C)=f_{AB}{}^C(t_C)^c{}_{f}\un e_c\otimes  f^f.
\eeqn
So if $v_E$ is a morphism, then we have the matrix equation
\beqn
\left[t_A,t_B\right]{}^a{}_b&=&f_{AB}{}^C(t_C){}^a{}_b.
\eeqn
In the special case where $E$ is the bundle $L$, we have the  result $v_L(\un t_A)=f_{AC}{}^B \un t_B\otimes t^C$, the adjoint representation.
} of the map $v_L:L\to End(L)$ as
\beq\label{repconvertbracket}
[\omega(\un\mX_V),\un\mu]_L=v_L\circ\omega(\un\mX_V)(\un\mu).
\eeq
Here, we regard $v_L\circ\omega(\un\mX)\in End(L)$. 
From eq. \eqref{defRomega}, the first term in \eqref{LbdlderivLsplit} is equal to $-R^\omega(\un\mX_H,\iota(\un\mu))$, and we introduce the notation\footnote{It is easy to see that $R^\omega(\un\mX_V,\un\mY_V)=\un 0$  which we can interpret as the statement that the Lie algebra has no curvature. We also saw that $R^\omega(\un\mX_H,\un\mY_H)$ is related to curvature, and so $R^\omega(\un\mX_H,\un\mY_V)$ is the only remaining component, and makes its appearance in \eqref{defcovLderiv}.}
\beq\label{defcovLderiv}
\omega([\un\mX_H,\iota(\un\mu)]_A)=R^\omega(\un\mX_H,\iota(\un\mu))
\equiv -\nabla^L_{\un\mX_H}\un\mu.
\eeq
Note that we have written the section $\un\mX_H\in\Gamma(A)$ as a subscript to $\nabla^L$ to denote a directional derivative; since this section is horizontal, we could equally well have written $\rho(\un\mX_H)\in\Gamma(TM)$ to emphasize that the derivative is along a tangent to $M$ in the spirit of eq. \eqref{rhoLeib}. 
We use the $\nabla$ notation here because it depends on the connection and, as we will see later, in gauge theories it will become the gauge covariant derivative along $\rho(\un\mX_H)$. It is also natural to write $\nabla^L\un\mu\in\Gamma(A^*\times L)$, with $\nabla^L\un\mu(\un\mX)\equiv \nabla^L_{\un\mX_H}\un\mu$. 

To summarize, we have that the derivation on $L$, defined by the Lie bracket on $A$ as in eq. \eqref{Lderrep}, reads
\beq\label{LbdlderivL2}
\phi_L(\un\mX)(\un\mu)=
\nabla^L_{\un\mX_H}\un\mu-v_L\circ\omega(\un\mX_V)(\un\mu)
\eeq
and so we have split $\phi_L$ into a horizontal and vertical part, the horizontal part having an interpretation as a covariant derivative, the vertical being an endomorphism. Given the definition of $\hatd$ in \eqref{defphiDE}, applied to $L$, 
\beq
\hatd\un\mu(\un\mX)=\phi_L(\un\mX)(\un\mu),
\eeq
we can interpret the above result as
\beq
\hatd\un\mu(\un\mX_H)=\nabla^L_{\un\mX_H}\un\mu,\qquad
\hatd\un\mu(\un\mX_V) \equiv s\un\mu(\un\mX)=-v_L\circ\omega(\un\mX_V)(\un\mu).
\eeq
Here we have introduced the notation $s\un\mu$ to denote the vertical part of $\hatd\un\mu$. 
Later we will generalize these expressions from $\un\mu\in\Gamma(L)$ to general $E$-valued $n$-forms. 

Before moving on, let us note that $\nabla^L$ has a curvature which we define in the usual way, as a map ${\cal R}^L:H\times H\times L\to L$,
\beqn\label{Lcurvdef}
{\cal R}^L(\un\mX_H,\un\mY_H)(\un\mu)&=& 
\nabla^L_{\un\mX_H}(\nabla^L_{\un\mY_H}\un\mu)
-\nabla^L_{\un\mY_H}(\nabla^L_{\un\mX_H}\un\mu)
-\nabla^L_{[\un\mX_H,\un\mY_H]_H}\un\mu,
\eeqn
which is linear in $\un\mX_H,\un\mY_H$ with respect to multiplication by functions on $M$. The notation $[\un\mX_H,\un\mY_H]_H$ refers to the horizontal part of the Lie bracket. We compute, making repeated use of \eqref{defcovLderiv},
\beqn
{\cal R}^L(\un\mX_H,\un\mY_H)(\un\mu)
&=&
-\nabla^L_{\un\mX_H}(\omega([\un\mY_H,\iota(\un\mu)]_A))
+\nabla^L_{\un\mY_H}(\omega([\un\mX_H,\iota(\un\mu)]_A))
+\omega([[\un\mX_H,\un\mY_H]_H,\iota(\un\mu)]_A)
\nonumber\\
&=&
\omega([\un\mX_H,\iota\circ\omega([\un\mY_H,\iota(\un\mu)]_A)]_A)
-\omega([\un\mY_H,\iota\circ\omega([\un\mX_H,\iota(\un\mu)]_A)]_A)
\nonumber\\&&
+\omega([[\un\mX_H,\un\mY_H]_A,\iota(\un\mu)]_A)
-\omega([[\un\mX_H,\un\mY_H]_V,\iota(\un\mu)]_A)
\nonumber\\
&=&
-\omega\Big([\un\mX_H,[\un\mY_H,\iota(\un\mu)]_A]_A
+[\un\mY_H,[\iota(\un\mu),\un\mX_H]_A]_A
+[\iota(\un\mu),[\un\mX_H,\un\mY_H]_A]_A\Big)
\nonumber\\
&&
-R^\omega([\un\mX_H,\un\mY_H]_V,\iota(\un\mu))
-[\omega([\un\mX_H,\un\mY_H]_V),\un\mu]_L)
\nonumber\\
&=&
-(v_L\circ\omega([\un\mX_H,\un\mY_H]_V))(\un\mu)
\nonumber\\
&=&
v_L(\Omega(\un\mX_H,\un\mY_H))(\un\mu),
\label{Lnablacurv}
\eeqn
where we used Jacobi, the fact that $R^\omega(\un\mX_V,\un\mY_V)=\un 0$, eq. \eqref{repconvertbracket} and introduced the notation $[\un\mX_H,\un\mY_H]_V$ to refer to the vertical part of the Lie bracket. Referring back to eq. \eqref{Russianformula}, we see that ${\cal R}^L$ involves just the previous notions of curvature; the presence of the map $v_L$ simply expresses the curvature as an endomorphism on $L$.

\subsubsection{The map $\phi_E$ for general vector bundles}\label{sec:phiE}

Returning to the general case of $\secE_0\in\Gamma(E)$, the Lie bracket can no longer be used to define $\phi_E$, but we propose that $\phi_E$ must be written in the same form as \eqref{LbdlderivL2},
\beq\label{Erep}
\phi_E(\un\mX)(\secE_0)=(\hatd\secE_0)(\un\mX)=\nabla^E_{\un\mX_H}\secE_0-v_E\circ\omega(\un\mX_V)(\secE_0)=\nabla^E_{\un\mX_H}\secE_0+s\secE_0(\un\mX_V).
\eeq
In the context of gauge theories, we will interpret $\nabla^E$ as the induced connection on the associated bundle $E$ and $\nabla^E_{\un\mX_H}\secE_0$ then as the covariant derivative of $\secE_0$ along $\rho(\un\mX_H)$.  
We must check that the given $\phi_E$ is in fact a morphism, which as shown will be equivalent to the nilpotency of $\hatd$; using \eqref{Erep}, we find
\beqn
[\phi_E(\un\mX),\phi_E(\un\mY)](\secE_0)
-\phi_E([\un\mX,\un\mY]_A)(\secE_0)
&=&
\phi_E(\un\mX)(\phi_E(\un\mY)(\secE_0))
-\phi_E(\un\mY)(\phi_E(\un\mX)(\secE_0))
-\phi_E([\un\mX,\un\mY]_A)(\secE_0)
\nonumber\\
&=&v_E(R^\omega(\un\mX,\un\mY))(\secE_0) + {\cal R}^E(\un\mX_H,\un\mY_H)(\secE_0)
\nonumber
\\&&
+\nabla^E_{\un\mY_H}(v_E\circ\omega(\un\mX)(\secE_0))
-v_E\circ\omega(\un\mX)(\nabla^E_{\un\mY_H}\secE_0)
\nonumber
\\&&
-\nabla^E_{\un\mX_H}(v_E\circ\omega(\un\mY)(\secE_0))
+v_E\circ\omega(\un\mY)(\nabla^E_{\un\mX_H}\secE_0),\label{phimor}
\eeqn
where we define the curvature of $\nabla^E$ analogously to \eqref{Lcurvdef},
\beq\label{Ecurvdef}
{\cal R}^E(\un\mX_H,\un\mY_H)(\secE_0)\equiv
\nabla^E_{\un\mX_H}(\nabla^E_{\un\mY_H}\secE_0)
-\nabla^E_{\un\mY_H}(\nabla^E_{\un\mX_H}\secE_0)
-\nabla^E_{[\un\mX_H,\un\mY_H]_H}\secE_0.
\eeq
The map $\phi_E$ under scrutiny is a morphism if and only if \eqref{phimor} vanishes.
It is simplest to understand \eqref{phimor} by considering separately horizontal and vertical sections. 
First note that if $\un\mX,\un\mY$ are both vertical this vanishes identically. 
If we take $\un\mX$ to be horizontal and $\un\mY=\iota(\un\mu)$, then, using \eqref{defcovLderiv}, we obtain
\beqn
[\phi_E(\un\mX_H),\phi_E(\iota(\un\mu))](\secE_0)-\phi_E([\un\mX_H,\iota(\un\mu)]_A)(\secE_0)
=
-v_E(\nabla^L_{\un\mX_H}\un\mu)(\secE_0) 
+\nabla^E_{\un\mX_H}(v_E(\un\mu)(\secE_0))
-v_E(\un\mu)(\nabla^E_{\un\mX_H}\secE_0).
\eeqn
The vanishing of this equation is essentially a compatibility condition between $\nabla^L$ and $\nabla^E$.
The remaining part of \eqref{phimor} has $\un\mX$ and $\un\mY$  both horizontal, and reads
\beqn
[\phi_E(\un\mX_H),\phi_E(\un\mY_H)](\secE_0)-\phi_E([\un\mX_H,\un\mY_H]_A)(\secE_0)
&=&v_E(R^\omega(\un\mX_H,\un\mY_H))(\secE_0) 
+{\cal R}^E(\un\mX_H,\un\mY_H)(\secE_0)
\\
&=&v_E\circ\omega([\un\mX_H,\un\mY_H]_A)(\secE_0) 
+{\cal R}^E(\un\mX_H,\un\mY_H)(\secE_0).
\eeqn
Thus the statement that $\phi_E$ is a morphism can be understood as the condition that ``curvature is curvature''. The curvature \eqref{Ecurvdef} acts on sections of $E$ as
\beq\label{Enablacurv}
{\cal R}^E(\un\mX_H,\un\mY_H)(\secE_0)=-v_E\circ\omega([\un\mX_H,\un\mY_H]_A)(\secE_0)
\eeq
which is of precisely the same form as \eqref{Lnablacurv}.
Recalling the result \eqref{OmegaRomega}, we can alternatively write this as
\beq\label{EnablacurvOmega}
{\cal R}^E(\un\mX_H,\un\mY_H)(\secE_0)=v_E(\Omega(\un\mX,\un\mY))(\secE_0).
\eeq
This is consistent with $\nabla^E$ being an induced connection, and furthermore shows that each notion of curvature that we have encountered is ultimately the same, geometrically associated to a non-trivial lift of $TM$ into $A$.

\subsubsection{The action of $\hatd$ on $E$-valued extended forms}\label{sec:hatdgenact}

We have given above, eq. \eqref{Erep}, the essential formula for derivations on an associated vector bundle $E$. The Koszul formula \eqref{koszulE} instructs how to extend the action of $\hatd$ on sections $\un\psi_n$ of $\wedge^nA^*\times E$, and furthermore, this may be split into vertical and horizontal pieces which we write as
\beq\label{dpsi}
\hatd\secE_n=(\nabla^{\wedge^nA^*\times E}\secE_n)+s\secE_n.
\eeq
%which generalizes \eqref{Erep}.
The two pieces of this equation may be written more explicitly as
\beqn\label{gencovder}
(\nabla^{\wedge^nA^*\times E}\secE_n)(\un \mX^1,...,\un\mX^{n+1})&=&
\sum_{j=1}^{n+1}(-1)^{j+1}(\nabla^{\wedge^nA^*\times E}_{\un\mX^j_H}\secE_n)(\un \mX^1,...,\widehat{\un\mX^j},...,\un\mX^{n+1}),
\\
(s\secE_n)(\un \mX^1,...,\un\mX^{n+1})&=&-\sum_{j=1}^{n+1}(-1)^{j+1}
v_E\circ\omega(\un\mX^j_V)(\secE_n)(\un \mX^1,...,\widehat{\un\mX^j},...,\un\mX^{n+1})
\label{genbrsttrans}\\
&&+\sum_{j<k}(-1)^{j+k}\secE_n([\un\mX^j_V,\un\mX^k_V],\un\mX^1,...,\widehat{\un\mX^j},...,\widehat{\un\mX^k},..,\un\mX^{n+1}).
\nonumber
\eeqn
The induced connection $\nabla^{\wedge^nA^*\times E}$ can be systematically constructed via Leibniz and linearity. This is explained for the case $n=1$ in Appendix \ref{app:LAmaps}, where we show how the compact result \eqref{dpsi} is obtained from the Koszul formula.
This completes the horizontal-vertical split for arbitrary $E$-valued extended forms, and in particular defines the vertical part $s$ generally, which is a key result for physical applications.

\subsection{The BRST interpretation}\label{sec:ghost}

So far, we have discussed the mathematics of transitive Lie algebroids. In the following section, we will describe a special case, the Atiyah Lie algebroids, which are associated with principal bundles and will be of direct relevance to gauge theories. Let us pause here though to explain how the Lie algebroids can be used in physical theories emphasizing some of the important structure that we have already established. 

In physics, fields will correspond either to gauge fields (which we will extract below) or to sections of associated bundles. We have understood how $\hatd$ implements (through the horizontal parts) a connection $\nabla$ on these bundles. As such the horizontal part tells us how to move sections of associated bundles along $M$. But we have also seen that $\hatd$ contains more, through the vertical part, which we referred to as $s$ above. That is, $\hatd$ combines displacements on the manifold $M$ with those associated with the bundle $L$. 
In fact, the vertical part of $\hatd$, which we called $s$ above, should be interpreted as the BRST transformation. This comes about because of the way this operator acts on sections of associated bundles, which is exactly the same as the BRST operator $s$. Although at this abstract stage this is an identification, we will show concretely in the following that such an identification is well justified, for we will reproduce the known BRST transformations from the vertical part of $\hat d$.

For sections of $E$, we wrote previously in \eqref{Erep},
\beq\label{brstsecE}
s\secE(\un\mX)=\hatd\secE(\un\mX_V)=-v_E\circ\omega(\un\mX_V)(\secE)\equiv -\ghost_E(\secE)(\un\mX_V),
\eeq 
and the nilpotency of $\hatd$ implies the nilpotency of $s$. Let us recall that $\omega:A\to L$, so $\omega(\un\mX_V)\in\Gamma(L)$. Further, since $v_E:L\to End(E)$, we can regard $\ghost_E=v_E\circ\omega\in\Gamma(A^*\times End(E))$, and so $v_E\circ\omega(\un\mX_V)\in End(E)$ acts as matrix multiplication on the components of $\secE$ in a basis for $E$. Equivalently, we can say that $v_E\circ\omega$ can be regarded as an $End(E)$-valued (that is, matrix-valued) vertical 1-form.

As already observed, the form of \eqref{brstsecE} is precisely that of a BRST transformation of the field $\secE$, with $s$ playing the role of the BRST generator, and $\ghost_E=v_E\circ\omega$ playing the role of the \hlt{ghost} field. In the usual physics (BRST) discussions, as for instance reviewed in \cite{Nemeschansky:1987xb, Henneaux:1989jq, Henneaux:1992ig, Becchi:1996yh, VanHolten:2001nj, Fuster:2005eg}, ghost fields are said to be Grassmann-valued. The important aspect of this representation is that they are anti-commuting. Here, the role of the Grassmann algebra is played by the vertical exterior algebra, which shares the anti-commuting property through the wedge product of forms. 
What we have here is an invariant way of describing this geometric structure. In the following section, we will discuss the specific case of Atiyah Lie algebroids which correspond more directly to gauge theories (as there is a group action involved), and this same BRST structure will carry over unchanged. In Section \ref{Sec4}, we will introduce local trivializations, which will allow us to make even more direct contact with physics notions. We will see then that the connection reform $\omega$ can be thought of as having two aspects: first, its components in a given basis will be interpreted as the components of the ghost field, and second, it will carry the information of the connection on the Lie algebroid through the basis form.
Thus the Lie algebroid formalism for gauge theories will have the remarkable property that the Grassmann nature of BRST is geometrized. 

\section{Atiyah Lie Algebroids}\label{sec:atiyah}

We now come to the central example of Lie algebroids for physics: the \hlt{Atiyah Lie algebroid}. The traditional account of the geometric formulation of gauge theories \cite{10.1143/PTP.16.537, Lubkin:1963zz, PhysRevD.12.3845, RevModPhys.52.175} involves equivariant connections on principal bundles. It has been previously suggested \cite{Fournel:2012uv, Jordan:2014uza, Attard:2019pvw} that that formalism may be replaced by considering connections on Atiyah Lie algebroids. As we will review, the passage from (the tangent bundle of) a principal bundle to the corresponding Atiyah Lie algebroid involves taking the quotient by the right action of the group. It is this right action that is a redundancy in the principal bundle formulation and in the required equivariance of the principal connections. 

In this section, we will provide a review of the geometry of principal bundles and their principal connections, and explain the effect of the quotient.  Our presentation, although perhaps technical and certainly not novel (e.g., see again \cite{mackenzie_1987, mackenzie_2005}), is not meant to be entirely rigorous, but instead to give an introduction to how an Atiyah Lie algebroid can be constructed, and why it is the right bundle to fully geometrize gauge theories. In particular, we will emphasize that through taking the quotient of the right action, we obtain a description of the physics of gauge theories involving only bundles over space-time itself and the gauge connections become simply Ehresmann connections without an equivariance requirement.

In the previous section, we have considered transitive Lie algebroids in general, and have noted that the exterior derivative on the Lie algebroid may be interpreted to contain, in its vertical part, the BRST operator. Consequently, as we pass to Atiyah Lie algebroids with their proposed physical interpretation, the usual physics notion of BRST also acquires a geometric interpretation, and this interpretation is present regardless of details of dynamics (that is, it is off-shell).
Although, as we mentioned, Lie algebroids have been advocated before as alternative formulations of gauge theory, we do not believe that the relationship with BRST has been previously made. A relation of BRST to principal bundles was made long ago \cite{Neeman:1979cvl, ThierryMieg:1979xe, ThierryMieg:1979kh, Baulieu:1981sb, ThierryMieg:1982un, ThierryMieg:1987um}, and so one might regard the presence of BRST in the Lie algebroid formalism as a natural consequence. Making this step explicit has nonetheless important consequences, as we will unravel in this section. A reader familiar with the mathematics of principal bundles  and principal connections may wish to skip to the end of the following subsection.  

%This similarity however is only skin deep, and in this section and particularly in Section \ref{Sec4}, we will explain in detail how the BRST ghosts are encoded in the geometry. 

\subsection{Principal bundle}\label{sec:princbdl}

Given a $d$-dimensional manifold $M$, with an atlas $\{U_i\}$ for which each $U_i$ is homeomorphic to an open set $V_i$ in $\mathbb{R}^d$, $\phi_i:U_i\to V_i$, the transition functions of the manifold are defined on $U_i\cap U_j$, 
\beq
\phi_{ij}=\phi_i\circ\phi_j^{-1},
\eeq
with the respective maps restricted to $U_i\cap U_j$. This allows to describe a manifold locally, and we want to do the same for principal bundles.

A fibre bundle $\pi:E\to M$ can be described as a short exact sequence
\beq\label{fibrebdlshortExactSeq}
\begin{tikzcd}
0
\arrow{r} 
& 
F
\arrow{r}
& 
E
\arrow{r}{\pi} 
& 
M
\arrow{r} 
&
0
\end{tikzcd}
\eeq
where $F$ is the fibre, which might be a vector space or  a group. Locally, within an atlas on $M$, a fibre bundle $\pi:E\to M$ can be described as $(x,f)$ where $x\in U_i$ and $f\in F$.\footnote{We alternately think of $x$ as a point in $U_i$, or as its coordinates $\phi_i(x)$, which is a point in $V_i$. Also, we consider the inclusion of $F$ in $E$ trivial so we consider them indistinguishably.}  An example of a vector bundle that we get for free for any $M$ is the tangent bundle $TM$ where the fibre at a point $x$ is the tangent plane $T_xM$, and thus a point in $TM$ can be described locally as $(x,\un v_x)$ where $\un v_x\in T_xM$. 
We also get for free the cotangent bundle $T^*M$, which is also a vector bundle, the dual of $TM$; in this case the points in the fibres are 1-forms, which are maps from $T_xM\to \mathbb{R}$. 

What we mean by ``locally'' above is as follows. A \hlt{local trivialization} of the fibre bundle is an atlas for $M$ together with a homeomorphism $\varphi_i:\pi^{-1}(U_i)\to U_i\times F$ for all $x\in U_i$ and all $i$, such that $\pi\circ\varphi_i^{-1}(x,f)=x$ for all $f\in F$. So locally, $E$ looks like $U\times F$:
\beq\label{fibrebdlshortExactSeq}
\begin{tikzcd}
& 
& 
E
\arrow{r}{\pi} 
\arrow{dd}{\varphi}
& 
M
\arrow{r} 
&
0
\\
0
\arrow{r} 
& 
F
\arrow{ru}
\arrow{rd}
& 
& 
&
\\
& 
& 
U\times F
\arrow{r} 
& 
U
\arrow{r} 
\arrow{uu}
&
0
\end{tikzcd}.
\eeq

The homeomorphisms define transition functions on overlaps $U_i\cap U_j$ for the bundle via
\beq
\varphi_{ij}=\varphi_i\circ\varphi_j^{-1}:(U_i\cap U_j)\times F\to (U_i\cap U_j)\times F.
\eeq
At least in the case where $F$ is a vector space of dimension $k$, we have
\beq
\varphi_i\circ\varphi_j^{-1}(x,\un v_x)=(x,g_{ij}(x)\un v_x),\quad \forall x\in U_i\cap U_j,\ \un v_x\in F_x,
\eeq
where $g_{ij}: U_i \cap U_j \to GL(k,\mathbb{R})$, often restricted to lie in a subgroup. This is a \u{C}ech cocycle, in that we require
\beq
g_{ii}(x)=Id,\qquad g_{ij}(x)g_{jk}(x)g_{ki}(x)=Id,\ \ for\ x\in U_i\cap U_j\cap U_k
\eeq
A (local) section of $E$ is a continuous map $s:U_i\to E$ such that $\pi\circ s(x)=x,\forall x\in U_i$.

A \hlt{principal bundle} $\pi: P\to M$ is a fibre bundle whose fibres are a Lie group $G$ and is endowed with a continuous right action $R:P\times G\to P$, such that $R_h:P\to P$ for $h\in G$. $R$ acts freely: the stabilizer of every point is trivial. In the following we will write $R_h(p)=p\cdot h$. 
For the principal bundle, we can write a short exact sequence
\beq\label{principalshortExactSeq}
\begin{tikzcd}
0 \arrow{r} & G \arrow{r} & P \arrow{r}{\pi} & M \arrow{r} & 0
\end{tikzcd}
\eeq
and essentially all we said about vector bundles carries over here. 
A local trivialization (or ``choice of gauge") of $P$ is an atlas for $M$ together with a map $T_i:  \pi^{-1}(U_i)\to U_i\times G$ given by $T_i(p)=(\pi(p),t_i(p))$ where $t_i:\pi^{-1}(U_i)\to G$ is equivariant under right action in $P$, $t_i(p\cdot h)=t_i(p)h$, which implies
\beq
T_i(p\cdot h)=(\pi(p\cdot h),t_i(p\cdot h))=(\pi(p),t_i(p)h),
\eeq
where we used  that $\pi$ is invariant under the  right action of the group, $\pi(p\cdot h)=\pi(p)$.
So under these definitions, the right action of the group acts along the fibres and acts on the group itself as {right multiplication} via $t_i$.\footnote{A local trivialization $T_i$ is equivalent to specifying a local section $\sigma_i: U_i\to \pi^{-1}(U_i)$ such that $\pi\circ\sigma=Id_{U_i}$. We have simply $\sigma_i(x)=T_i^{-1}(x,e)$ and/or $T_i(\sigma_i(x)\cdot h)=(x,h)$, where $x\in U_i$, $h\in G$ and $e$ is the identity in $G$.}

On overlaps, we take the transition functions, as we wrote above, to be given by {left multiplication}, 
\beq
(\pi(p),t_i(p))=T_i\circ T^{-1}_j(\pi(p),t_j(p))=(\pi(p),\psi_{ij}(p)t_j(p))
,\qquad \psi_{ij}(p)=t_i(p) t_j(p)^{-1}.
\eeq
Thus the gauge transformations are understood to be associated with left multiplication, and the $\psi_{ij}$ are invariant under the right action of $G$
\beq
\psi_{ij}(p\cdot h)=t_i(p\cdot h) t_j(p\cdot h)^{-1}=t_i(p)hh^{-1} t_j(p)^{-1}=\psi_{ij}(p).
\eeq
 Hence it is conventional to write $\psi_{ij}(p)$ as $\psi_{ij}(x)$. This is already a good indication that, for physical applications, the right action of the group is a redundancy, i.e., transition functions are invariant.

Since we are interested in connections on principal bundles, the main object of interest is not $P$ itself but its tangent bundle $\hat\pi:TP\to P$. The bundle projection $\pi:P\to M$ pushes forward to a map $\pi_*:TP\to TM$, and $\pi_*$ will have a kernel called the vertical subspace $V_pP\subset T_pP$. A section $\un\mX\in \Gamma(TP)$ is then vertical if $\un\mX_p\in V_pP$ for all points $p$. As we recalled above, the right action of the group acts vertically on $P\to M$. Given $\un\mX_p\in T_pP$, we can generate a section by pushing forward the right action. This is a special class of sections, which are automatically right-invariant. That is
\beq
R_{*h}:T_pP\to T_{p\cdot h}P,
\eeq
and the map $R_{*h}$ is a morphism of the Lie bracket on $TP$,
\beq
R_{*h}([\un \mX,\un \mY]_p)=[R_{*h}(\un \mX_p),R_{*h}(\un \mY_p)]_{p\cdot h},\qquad\forall \un\mX,\un\mY\in\Gamma(TP), \forall p\in P.
\eeq
Then, a section $\un \mX\in\Gamma(TP)$ is right-invariant if for all $p\in P$ it satisfies\footnote{\label{foot:localright} In a local trivialization, if we write $\pi_*(\un\cX_{p})=\un X_{\pi(p)}\in T_{\pi(p)}M$, we have $\un\gamma_{p}=\un\cX_{p}-\pi_*^{-1}(\un X_{\pi(p)})\in \ker(\pi_*)$. We then regard locally $\un \mX_p\simeq (\un X_{\pi(p)},\un \gamma_p)$. 
 Right invariance therefore means that \[R_{*h}(\un X_{\pi(p)},\un\gamma_{p})\equiv (\un X_{\pi(p\cdot h)},R_{*h}(\un\gamma_{p}))=(\un X_{\pi(p)},\un \gamma_{p\cdot h}).\]
}
\beq
R_{*h}(\un\cX_p)=\un \cX_{p\cdot h}.
\eeq

The vertical vector space, at each point $p$, is isomorphic to the Lie algebra $\mg$ of the group $G$. That is, there is a map $j_p:\mg\to T_pP$ given by 
\beq
j_p(\un\gamma)={d\over dt}(p\cdot e^{t\un\gamma})\big|_{t=0},\quad \un\gamma\in \mg,
\eeq
which satisfies $\pi_* j_p(\un\gamma)=0$ for all $p\in P$ and thus $j(\un\gamma)\in \Gamma(VP)$ for all $\un\gamma\in \mg$. The right action on $V$ gives
\beq\label{adiota}
R_{*h}j_p(\un\gamma)=j_{p\cdot h}(Ad_{h^{-1}} \un\gamma), \qquad \text{with}\qquad Ad_h(\un\gamma)={d\over dt}(h e^{t\un\gamma} h^{-1})\big|_{t=0}.
\eeq
Thus the vertical sub-bundle $VP\subset TP$ is an invariant distribution under the group right action. 

A \hlt{principal connection} on $P$ can be defined as  a $G$-equivariant 1-form on $P$  valued in the Lie algebra $\mg$, i.e.,  $\omega\in\Gamma(T^*P\times\mg)$. 
For a right-invariant section $\un\mX\in\Gamma(TP)$, a principal connection $\omega$ satisfies
\beq\label{Rom1}
(R^*_{h}\omega_{p\cdot h})(\un\mX_p)
=\omega_{p\cdot h}( R_{*h}\un\mX_{p})
=Ad_{h^{-1}}(\omega_{p}( \un\mX_{p})),
\eeq
or more briefly $Ad_h(R^*_{h}\omega)=\omega$. Here we used the pull back $R^*_h:T^*_{p\cdot h}P\to T^*_{p}P$.

Another notion of connection on $P$ is introduced \`a la Ehresmann specifying a distribution $HP$ on $TP$,  complementary to $VP$. This connection is associated to a principal connection $\omega$ only when the distribution $HP$ is right invariant. If this is the case, then
\beq
\omega(\un\mX)=0\quad \text{if} \quad \un\mX\in \Gamma(HP) \qquad \text{and} \qquad \omega(j(\un\gamma))=-\un\gamma,
\eeq
where the minus sign conforms with our conventions. Using \eqref{Rom1} for $j(\un\gamma)$ we simply have
\beq
(R^*_{h}\omega_{p\cdot h})(j_p(\un\gamma))
=-Ad_{h^{-1}}(\un\gamma).
\eeq

At this point, this construction starts to strongly resemble the more general one discussed in the previous section for Lie algebroids. Nonetheless, the results expressed here apply only to right invariant sections in $\Gamma(TP)$. Moreover, it is important to observe that $TP$ is a bundle over $P$, whereas the general Lie algebroids discussed previously are bundles over $M$. Note that the set of these sections is closed under Lie bracket, since 
\beq
[\un\mX_{p\cdot h},\un\mY_{p\cdot h}]=[R_{*h}(\un\mX_p),R_{*h}(\un\mY_p)]=R_{*h}([\un\mX,\un\mY]_p)
=[\un\mX,\un\mY]_{p\cdot h}.
\eeq
So the set of right-invariant sections  on $P$, denoted $\Gamma_G(TP)$, is a closed subset of $\Gamma(TP)$. While $TP$ can be identified with  $\Gamma(TP)$, one may ask what bundle we can identify with $\Gamma_G(TP)$. The answer is the bundle $TP/G$, that is, the quotient of $TP$ by the right action of the group. This means that the set $\Gamma(TP/G)$ is in one to one correspondence with $\Gamma_G(TP)$.  Furthermore, as we will show shortly, $TP/G$ is a vector bundle over $M$, and the principal connection on $P$ descends to a connection on $TP/G$, which is the so-called Atiyah Lie algebroid.

\subsection{Constructing the Atiyah Lie algebroid $TP/G$}\label{sec:atiyahconstr}

Given a principal bundle $\pi:P\to M$ with a structure group $G$, there is a corresponding Atiyah Lie algebroid $TP/G$ that we may systematically construct. 
To define $TP/G$, recall that a point in $TP$ is locally described by $(p,\un v_p)$ with $\un v_p\in T_pP$, and the group acts (freely and transitively) on the right as $(p,\un v_p)\mapsto (p\cdot h,R_{*h}(\un v_p))$. 
Thus $TP/G$ is defined by identifying 
\beq\label{eq}
(p,\un v_p)\sim (p\cdot h,R_{*h}(\un v_p)),\qquad \forall h\in G.
\eeq
The resulting equivalence classes can be thought of in two ways, the first leading to an interpretation in terms of a vector bundle over $M$, and the second leading to right-invariant sections over $P$. 
Referring to the local description in footnote \ref{foot:localright}, we know that we can regard locally $\un v_p\simeq (\un X_{\pi(p)},\un \gamma_p)$. For each point in $TP/G$, we can simply select a convenient representative of the equivalence class in eq.  \eqref{eq} for which  $p\simeq (\pi(p),e)$ and thus $\un v_p\simeq(\un X_{\pi(p)},\un\gamma_{(\pi(p),e)})$. Then denoting $\pi(p)=x$, the point in $TP/G$ has a local description $(p,\un v_p)\simeq ((x,e),(\un X_x,\un\gamma_{(x,e)}))$, or more simply  $(x,\un X_x,\un\gamma_{(x,e)})$ with $\un\gamma_{(x,e)}$ recognized as an element of the Lie algebra $\mg$ associated to $G$. This implies that $TP/G$ is a vector bundle over $M$ rather than $P$, with the fibre over $x$ described locally by $(\un X_x,\un\gamma_{(x,e)})$, consisting of a tangent vector and a Lie algebra element. Thus, $TP/G$ is a vector bundle of rank $d+\dim G$ over $M$.

Alternatively, starting from a point $(p,\un v_p)\in TP$, we can regard \eqref{eq} as simply defining a right-invariant section over $P$: since $R_h$ acts transitively, the equivalence simply defines a section for which $\un v_{p\cdot h}$ is given by $R_{*h}\un v_p$ for all $h$ (and thus all points in $P$). 
This interpretation of the quotient makes direct contact with $\Gamma_G(TP)$, the subset $\Gamma_G(TP)\subset \Gamma(TP)$ of right invariant sections over $P$, as previously claimed. 

The interpretation of $TP/G$ as a vector bundle over $M$ can be taken further noticing that
\[
\pi_*:TP/G\to TM,
\]
in the local description, has a kernel consisting of vectors of the form $(\un 0,\un\gamma_{(x,e)})$. So we see that at each point $x$, $\ker\pi_*\simeq \mg$, as expected. The union over all points $x\in M$ gives a bundle of Lie algebras which can be identified with the
 \hlt{adjoint bundle} $L_G\equiv P\times_{Ad_G}\mg$, which  is a vector bundle over $M$, associated to $P$, whose fibre is the Lie algebra of $G$.\footnote{Locally, we can understand $L_G$ as being the space of equivalence classes $((x,g),\un\mu_{x})\sim ((x,gh),ad_h\un\mu_{x})$, a representative of which can be taken to be $((x,e),ad_{g^{-1}}\un\mu_{x})$, which we associate to $(x,\un\nu_x)$ with $\un\nu_x\in \mg$.} The map $j$ introduced above, extended to all points $p\in P$, is then the morphism $\iota:L_{G}\to TP/G$, where the adjoint representation comes about thanks to \eqref{adiota}. We eventually have all of the structure necessary to interpret $TP/G$ as a Lie algebroid with $\pi_*$ being the anchor map and $L_G$ being the kernel of $\pi_*$ via $\iota$,
\beq\label{AtiyahLiealgebroidseq}
\begin{tikzcd}
0 \arrow{r} & L_G \arrow{r}{\iota} & TP/G \arrow{r}{\pi_*} & TM \arrow{r} & 0
\end{tikzcd}.
\eeq
This is the Atiyah Lie algebroid corresponding to the principal bundle $P$. While for abstract transitive Lie algebroids the bundle $L$ was introduced to define the short exact sequence \eqref{shortExactSeqGenLieAlgd}, in the context of Atiyah Lie algebroids the adjoint bundle $L_G$ acquires an important physical interpretation. As reviewed in \cite{RevModPhys.52.175}, the space of sections of $L_G$ is indeed the algebra of infinitesimal gauge transformations, and this active definition of gauge transformations is completely equivalent to the definition given in \cite{Atiyah:1978wi}, where a gauge transformation on $P$ is introduced as a vertical equivariant bundle isomorphism. So in this sense, the adjoint bundle $L_G$ is a necessary ingredient of any classical gauge theory. Notice eventually that, as again shown in \cite{RevModPhys.52.175}, these definitions are equivalent to the passive local interpretation of gauge transformations that we will give in Section \ref{Sec4}, once the concept of local trivialization of the Lie algebroid has been introduced.

An Ehresmann connection on this algebroid defines a complementary sub-bundle $H$ to $\ker \pi_*$ in $TP/G$, which is now automatically right invariant. A principal connection on $P$ passes through the quotient to an Ehresmann connection on $TP/G$. Thus, as far as physics is concerned, a connection on an Atiyah Lie algebroid serves as an appropriate geometric interpretation of a gauge field, the structure of which we will explore in detail in what follows. 

\subsection{The physics}\label{sec:physicsTPmodG}

The idea of formulating gauge theories geometrically on principal bundles was described originally in \cite{10.1143/PTP.16.537, Lubkin:1963zz, PhysRevD.12.3845, RevModPhys.52.175}. Later on, with the advent of BRST \cite{Becchi:1975nq, Tyutin:1975qk}, the question of how to encompass this  formalism on principal bundles arose.
In the '$80$s, Thierry-Mieg and collaborators \cite{Neeman:1979cvl, ThierryMieg:1979xe, ThierryMieg:1979kh, Baulieu:1981sb, ThierryMieg:1982un, ThierryMieg:1987um} observed that it is possible to identify the gauge ghost with the ``vertical part'' of an extended form, the principal connection, on the principal bundle $P$. 
We briefly review here the discussion in \cite{ThierryMieg:1979kh,ThierryMieg:1982un,ThierryMieg:1987um}. The idea is to choose coordinates along the fibres of $P$ (within a local trivialization) and write extended quantities (denoted by tilde) in $P$. Defining a section of $P$ from the base $M$ via the map $\Sigma: M\to P$ selects a gauge choice. Consider then an extended one form $\tilde A$ on $P$ valued in a Lie algebra. This extended one form is a principal connection if it is equivariant, as reviewed above. In this case, its pullback on $M$, $\Sigma^*(\tilde A)$, is the Yang-Mills one form in the gauge defined by $\Sigma$. However, $\tilde A$ contains more information encoded in its "vertical" part. This was recognized to be the Faddeev-Popov ghost \cite{Faddeev:1967fc}. Thierry-Mieg introduced an exterior derivative in $P$, $\tilde d$ and defined coordinates $x^\mu$ on $M$ and $y^a$ on the fibres of $P$.  The coordinates $y^a$ are related to the Lie algebra at each point. Then, the components $\tilde A^a$ of the Lie algebra-valued extended form $\tilde A$ are written as
\beq
\tilde A^a=A^a_\mu \tilde d x^\mu+A^a_b \tilde d y^b
\eeq
such that $c=A^a_b \tilde d y^b$ is the Faddeev-Popov ghost. Furthermore, the BRST operator was identified to be
\beq
s= \tilde d y^a \tilde{\un \pa}_a.
\eeq
Then, conditions like the horizontality of the curvature $2$-form (i.e., the Russian formula) followed upon imposing equivarance of the connection $\tilde{A}$, and gave rise to the known BRST transformations. Furthermore, the discussion was entirely local, since their description required coordinates to be introduced along the fibres. 
%Although there is nothing wrong in using local coordinates along the fibres, it is not clear how general the results are. 
One of the main achievements was the understanding of the Grassmann algebra geometrically as the algebra of vertical forms.

Subsequently, Bonora and Cotta-Ramusino remarked in \cite{Bonora:1982ve} that Thierry-Mieg's construction has a problem: a principal connection on $P$ is a one form in $T^*P$ valued in the Lie algebra. As we saw, $TP$ is not a bundle over $M$ but over $P$ instead. Components of forms in $T^*P$ are therefore functions on $P$. The ghost however is a function over $M$ valued in the Lie algebra. While for the gauge connection part of $\tilde{A}$ this issue is eliminated via the pullback $\Sigma^*(\tilde{A})$, the vertical part should not be straightforwardly related to the ghost, because the former is defined on $P$ while the latter on $M$. Bonora and Cotta-Ramusino realized that the ghost can be identified with the Maurer-Cartan form on the gauge group  ${\cal G}\equiv P\times_{Ad_G}G$, which  is a vector bundle over $M$, associated to $P$, whose fibre can be identified with the Lie group $G$. In this way the ghost becomes universal (it is associated to a canonical geometric structure), and both issues mentioned above are lifted. The horizontal and vertical complexes were introduced in \cite{Bonora:1982ve} separately as the de Rham complex (directly on $M$) with values in the Lie algebra, and the BRST complex in the space of connections. This comes about because, by construction, the complex on $P$ would have contained again extra spurious structure, for the same reason mentioned above.

The construction we  give in the present paper in terms of Atiyah Lie algebroids is of course closely related to these previous works, and can be considered as making explicit the global geometric structure. 
As we have discussed, an Atiyah Lie algebroid simplifies the geometry of gauge theories by directly quotienting out the right group action, yielding a vector bundle over $M$ with connection. The geometric content of the algebroid construction remains in one-to-one correspondence with the physical content of the principal bundle construction with principal connection. 
Then, all the results, including the ghost interpretation of \cite{Bonora:1982ve}, hold true within a local description of Atiyah Lie algebroids. Nevertheless, we think it is worth the effort to understand the structure of Atiyah Lie algebroids carefully, as it  implements by construction  the qualifications (such as the restriction of attention to right-invariant sections of $TP$) that are necessary in the principal bundle formalism. 

So in this sense one can interpret the Atiyah Lie algebroid as a minimal mathematical model underlying gauge theories. So far in the paper, we have described features of Lie algebroids without depending upon local realizations. In the next section, we will describe how to introduce a local trivialization of an Atiyah Lie algebroid along with an atlas for $M$, with the intent of making contact with the usual physical descriptions of gauge theories. We will see that there are some similarities to the local structures introduced in the previously cited works. However, we should note that there are important conceptual differences; for example, in the Atiyah Lie algebroid formalism, one achieves the complex $\Omega^\bullet(A,E)$, with $A=TP/G$ and $E$ the associated bundle of matter fields, which generalizes the local complex (interpreted here in terms of a trivialization of $A$) introduced in \cite{Bonora:1982ve}.

\section{Yang-Mills Gauge Theories}\label{Sec4}

In this section, our intent is to bring the abstract (but intrinsic) notation of the previous sections a little more down to earth. In particular, we will consider the local trivializations of the Lie algebroid and the vector bundles $E$, $L$ and $TM$. 
For a gauge theory application, one must make a choice of structure group $G$ and a selection of associated bundles. The physical fields then correspond to the connection as well as the sections of the associated bundles (i.e., the charged matter fields).
Each of the associated bundles $E$ has a left $G$ action in a representation $R$, whose dimension is the rank of $E$. This left action is the gauge transformation, which we will be able to define once the bundles are trivialized. A local trivialization will involve a choice of atlas for $M$, and a local choice of basis for the fibres of each bundle.  We will see that the extraction of the physical gauge field as well as the gauge ghosts is a little intricate: the gauge field in particular appears in several guises. We will generally use the notation $A=TP/G$ for the Atiyah Lie algebroid and we will write $L=L_G$, leaving the appellation $G$ implicit.  We begin the discussion with some general comments about bases of sections for the various vector bundles, which will be used in the local description.

Let us begin by recalling the properties of the various maps involved, $\rho: A\to TM$, $\iota: L\to A$, $\sigma: TM\to A$ and $\omega: A\to L$, with the properties:
\beqn
\rho \circ \sigma = Id_{TM}\label{rs},&\qquad &
\omega \circ \iota =- Id_{L},\\
\rho \circ \iota = 0,&\qquad&
\omega \circ \sigma = 0.\label{os}
\eeqn
We will choose specific sets of indices to label bases of sections of each vector bundle: we write $\{\un E_M\}$ ($M=1,\dots,\dim A$) as a basis for $A$, $\{\un \pa_\mu\}$ ($\mu=1,\dots,\dim M$) as a (coordinate) basis for $TM$ and $\{\un t_A\}$ ($A=1,\dots,\dim G$) as a basis for $L$ (consult Appendix \ref{sec:notation} for further details on conventions). We then have explicitly
\beqn
\rho(\un E_M)= \rho^\mu{}_M \un\pa_\mu,\qquad\qquad
\sigma(\un \pa_\mu)=\sigma^M{}_\mu \un E_M\\
\iota(\un t_A)=\iota^N{}_A\un E_N,\qquad\qquad
\omega(\un E_M)=\omega^A{}_M\un t_A.
\eeqn
Relationships (\ref{rs}-\ref{os}) read then
\beqn
\rho^\nu{}_M\sigma^M{}_\mu=\delta^\nu{}_\mu,\qquad&&
\omega^A{}_M\iota^M{}_B=-\delta^A{}_B,
\\
\rho^\mu{}_M\iota^M{}_A=0,
\qquad&&
\omega^A{}_M\sigma^M{}_\mu=0\label{omsi}.
\eeqn
Correspondingly, contractions $\sigma^M{}_\mu\rho^\mu{}_N$ and $\iota^M{}_A\omega^A{}_N$ are horizontal and vertical projectors, respectively, for sections of $A$. 
That is, for a section $\un\mX$ of $A$, we have
\beq
\un\cX=\cX^M\un E_M=\un\mX_H+\un\mX_V=\cX^M \sigma^N{}_\mu \rho^\mu{}_M \un E_N-\cX^M\iota^N{}_A \omega^A{}_M \un E_N.
\eeq
It is clear that we can change basis on $A$ at will,  
\beq\label{JE}
\un E_M=J^N{}_M\un E^\prime_N.
\eeq
Correspondingly, the components of $\un\mX$ transform as
\beq\label{JX}
\mX^{\prime M}= J^M{}_N\mX^N,
\eeq
such that $\un\mX$ is invariant, that is a tensor.
Given a basis $\un E_M$ for $A$, we have a dual basis for $A^*$, which we write as $E^M$ satisfying $E^M(\un E_N)=\delta^M{}_N$. 
In the next subsection, we will make use of such transformations $J^M{}_N$ to specify further adapted bases that block-diagonalize the projectors.

\subsection{The horizontal distribution and split bases}\label{sec:appsplit}

The vertical distribution $V\subset A$ is integrable, because it is an ideal with respect to Lie brackets in $A$. It has rank $\dim L$ and enjoys two properties: it is the image of the morphism $\iota$, and defines the kernel of the anchor $\rho$. The horizontal distribution $H\subset A$, complementary to $V$, is defined as the image of the map $\sigma$. We can consequently introduce in $A$ a separation of indices $M=(\alpha,\un A)$ with $\un A=(1,\dots,\dim G)$ such that the basis $\un E_{\un A}$ spans $V$ and $\alpha=(1,\dots,\dim M)$ such that $\un E_\alpha$ spans $H$. We will refer to this as a \hlt{split frame}. This will in general involve the use of \eqref{JE}, and will lead in particular to a block-diagonalization of the projectors referred to in the previous subsection. 
The maps $\iota$ and $\sigma$ can be written as
\beqn
\iota(\un t_A) &=&\iota^{\un A}{}_A\un E_{\un A}+\iota^{\alpha}{}_A\un E_{\alpha},\\
\sigma(\un\pa_\mu)&=&\sigma^\alpha{}_\mu\un E_\alpha+\sigma^{\un A}{}_\mu\un E_{\un A},
\eeqn
and requiring that their images are $V$ and $H$ respectively corresponds to setting
\beqn\label{iH}
\iota^{\alpha}{}_A=0,\qquad
\sigma^{\un A}{}_{\mu}=0.
\eeqn
Given that $V$ is the kernel of $\rho$, and $H$ the kernel of $\omega$, we have also
\beqn\label{rhoV}
\rho(\un E_{\un A})=0, \quad \Rightarrow \quad \rho^\mu{}_{\un A}=0,\qquad\qquad
\omega(\un E_\alpha)=0,\quad \Rightarrow \quad \omega^A{}_\alpha=0.
\eeqn
The  index structure  responsible for \eqref{iH} and \eqref{rhoV}  can be alternatively seen as a specific solution of the equations $\rho\circ\iota=0$ and $\omega\circ\sigma=0$. So in a split frame (at each point $x\in M$), the maps satisfy (\ref{iH}-\ref{rhoV}), but the remaining components, namely $\iota^{\un A}{}_A$, $\omega^A{}_{\un A}$, $\rho^\mu{}_\alpha$ and $\sigma^\alpha{}_\mu$ are unrestricted, apart from the relations\footnote{Eqs. \eqref{splitframeids} are just the split frame expressions of eqs. \eqref{rhosigid} and \eqref{omioid}. Note further that $\iota\circ\omega$ and $\sigma\circ\rho$ being projectors on $A$ to $V$ and $H$ respectively, we have $\iota\circ\omega\big|_V=-Id_V$ and $\sigma\circ\rho\big|_H=Id_H$. Thus, in a split frame we also have 
\beq
\sigma^\alpha{}_\mu\rho^\mu{}_\beta=\delta^\alpha{}_\beta, \qquad 
\iota^{\un A}{}_A\omega^A{}_{\un B}=-\delta^{\un A}{}_{\un B}.
\eeq
}
\beq\label{splitframeids}
%=\rho\circ\sigma
\rho^\nu{}_\alpha\sigma^\alpha{}_\mu=\delta^\nu{}_\mu, \qquad 
%=\omega\circ\iota
\omega^A{}_{\un A}\iota^{\un A}{}_B=-\delta^A{}_B.
\eeq
We can define the projected sections as above
\beq
\un\mX_H=\sigma\circ\rho(\un\mX)\in \Gamma(H), \qquad \un\mX_V=-\iota\circ\omega(\un\mX)\in \Gamma(V),
\eeq
and in the split basis we have
\beqn
\un\mX_H=\mX_H^\alpha\un E_\alpha
=\sigma^\alpha{}_\mu\rho^\mu{}_M \mX^M \un E_\alpha
=\sigma^\alpha{}_\mu\rho^\mu{}_\beta \mX_H^\beta \un E_\alpha,\hspace{0.7cm}
\\
\un\mX_V=\mX_V^{\un A}\un E_{\un A}
=-\iota^{\un A}{}_A\omega^A{}_{M} \mX^{M} \un E_{\un A}
=-\iota^{\un A}{}_A\omega^A{}_{\un B} \mX_V^{\un B} \un E_{\un A}.
\eeqn

Although such a split basis may always be chosen, the price to pay is to reduce the admissible basis transformations \eqref{JE} in $A$ to the ones preserving this structure,
\beqn\label{splitbasisV}
\un E_{\un A}=J^{\un B}{}_{\un A}\un E^\prime_{\un B} \quad \Rightarrow \quad J^\alpha{}_{\un A}=0,\\
\label{splitbasisH}
\un E_{\alpha}=J^{\beta}{}_{\alpha}\un E^\prime_{\beta} \quad \Rightarrow \quad J^{\un A}{}_{\alpha}=0.
\eeqn
Therefore the residual transformation matrices $J^N{}_M$ are block diagonal. Note that the dual basis for $A^*$ is also split, and we write the basis forms as $E^\alpha$ and $E^{\un A}$, with 
\beq\label{dualformorth}
E^\alpha(\un E_\beta)=\delta^\alpha{}_\beta,\quad 
E^\alpha(\un E_{\un A})=0,\quad 
E^{\un A}(\un E_\beta)=0,\quad
E^{\un A}(\un E_{\un B})=\delta^{\un A}{}_{\un B}.
\eeq

Here we have spoken in linear algebra terms, and so one should think in terms of this happening separately at each point $x\in M$. But given that specifying a connection on the Lie algebroid implies that globally $A=H\oplus V$, a split basis may be chosen at every point. In Section \ref{sec:apptriv}, we will make explicit the trivialization of each of the vector bundles and show how the $J^M{}_N$ matrices may be used in constructing transition functions.

\subsection{Lie brackets}\label{sec:brackets}

In addition, we have an algebraic structure on $A$. Given a basis of sections $\un E_M$, we define the rotation coefficients $C_{MN}{}^P$ as
\beq
\big[\un E_M,\un E_N\big]_A\equiv C_{MN}{}^P\un E_P.
\eeq
In a split basis, this can be decomposed into the collection
\beqn
\big[\un E_\alpha,\un E_\beta\big]_A&=& C_{\alpha\beta}{}^\gamma\un E_\gamma+C_{\alpha\beta}{}^{\un A}\un E_{\un A},\label{HHLie}\\
\big[\un E_\alpha,\un E_{\un A}\big]_A&=& C_{\alpha {\un A}}{}^{\un B}\un E_{\un B},\label{HVLie}\\
\big[\un E_{\un A},\un E_{\un B}\big]_A&=& C_{{\un A}{\un B}}{}^{\un C}\un E_{\un C},\label{VVLie}
\eeqn
where we made use of eqs. \eqref{HValgebraprops} to eliminate terms. Each of the rotation coefficients has a direct gauge theory interpretation, as we now discuss. Generally, the bracket satisfies the Leibniz property, eq. \eqref{rhoLeib}. Here, since $\rho(\un E_{\un A})=\un 0$, we have that \eqref{VVLie} is linear. Indeed, given two vertical sections $\un\mX_V,\un\mY_V\in\Gamma(A)$ we have
\beq
[\mX_V^{\un A}\un E_{\un A},\mY_V^{\un B}\un E_{\un B}]_A=\mX_V^{\un A}\mY_V^{\un B}[\un E_{\un A},\un E_{\un B}]_A=\mX_V^{\un A}\mY_V^{\un B}C_{\un A\un B}{}^{\un C}\un E_{\un C}.
\eeq
The rotation coefficients $C_{\un A\un B}{}^{\un C}$ are related to the Lie algebra structure constants of $L$ via the $\iota$ map,
\beq\label{rotstr}
C_{\un A\un B}{}^{\un C}\iota^{\un A}{}_A\iota^{\un B}{}_B=f_{AB}{}^C \iota^{\un C}{}_C,\qquad\qquad \big[\un t_A,\un t_B]_L=f_{AB}{}^C\un t_C,
\eeq
which follows from $\iota$ being a morphism of the Lie brackets.
This realizes the fact that at each point $x\in M$, there is a copy of the Lie algebra included into $A$.

This linearity does not extend to the rest of the rotation coefficients. For example, for a horizontal section $\un\mX_H\in\Gamma(A)$ and a vertical one $\iota(\un\mu)\in\Gamma(A)$ with $\un\mu\in\Gamma(L)$, applying \eqref{rhoLeib} to \eqref{HVLie} we obtain
\beq\label{HVbracket}
\big[\un\mX_H ,\iota(\un\mu)\big]_A= \mX_H^\alpha\Big(\iota^{\un A}{}_C\mu^C C_{\alpha {\un A}}{}^{\un B}+\rho(\un E_\alpha)(\iota^{\un B}{}_C\mu^C) \Big)\un E_{\un B}.
\eeq
Now, given eq. \eqref{defcovLderiv}, we have 
\beq\label{HVbracketnab}
[\un\mX_H,\iota(\un\mu)]_A=\iota(\nabla^L_{\un\mX_H}\un\mu).
\eeq
Generally, given a connection $\nabla^E$ (a directional derivative along a horizontal vector) on a vector bundle $E$ and a basis $\un e_a$ for $E$, we would define the connection coefficients in the given basis as
\beq\label{conncoeffsE}
\nabla^E_{\un\mX_H}\un e_a=A^b{}_a(\un\mX_H)\un e_b.
\eeq
What we will show in the next sub-section is that these connections coefficients are determined, for each associated bundle, by the same local data supplied by a trivialization of the Lie algebroid. This is a familiar result in gauge theory, but here it is a little more complicated, because we have a bundle of Lie algebras (rather than the {\it same} Lie algebra at each point in $M$). 

For the adjoint bundle specifically, we have
\beq\label{AinL}
\nabla^L_{\un\mX_H}\un t_A=A^B{}_A(\un\mX_H)\un t_B,
\eeq
so that
\beq
\iota(\nabla^L_{\un\mX_H}\un\mu) = 
\iota(\nabla^L_{\un\mX_H}(\mu^A\un t_A)) =
\mX_H^\alpha\Big(\rho(\un E_\alpha)(\mu^B)+\mu^AA_\alpha{}^B{}_A\Big)\iota^{\un B}{}_B\un E_{\un B},
\eeq
where we have written $A_\alpha{}^B{}_A\equiv A^B{}_A(\un E_\alpha)$.
Comparing to \eqref{HVbracket}, we see that 
\beq
 C_{\alpha {\un A}}{}^{\un B}\iota^{\un A}{}_C+\rho(\un E_\alpha)(\iota^{\un B}{}_C)  = \iota^{\un B}{}_B A_\alpha{}^B{}_C,
\eeq
and thus $\iota$, in its role as a map between $L$ and $V$, induces a non-linear relationship between the rotation coefficient and the connection coefficients. This equation has a geometric interpretation in terms of parallel transporting a vertical section horizontally in $A$: this action requires the adjoint bundle connection coefficient together with the horizontal derivative of the map $\iota$. Note that, given that the map $\iota$ restricted to $V$ is an isomorphism, one could set it to the identity in a local trivialization (discussed shortly). This is however achievable only in a particular chart, and thus, in order to establish global results, we will not make such an assumption.

Now using \eqref{omgcurvature2}, we see also that the vertical part of \eqref{HHLie} gives
\beq
C_{\alpha\beta}{}^{\un A}=\Omega^A(\un E_\alpha,\un E_\beta)\iota^{\un A}{}_A\equiv\iota^{\un A}{}_A\Omega^A_{\alpha\beta}
\eeq
which is linear and determined just by the curvature. As already noticed, this is the curvature of the distribution $H$ that makes it non-integrable.
Finally, supplying two horizontal sections $\sigma(\un X),\sigma(\un Y)\in \Gamma(A)$ with $\un X,\un Y\in \Gamma(TM)$, the horizontal part of \eqref{HHLie} evaluates to
\beqn
\big[\sigma(\un X),\sigma(\un Y)\big]_A^\alpha&=&
\sigma^\beta{}_\mu X^\mu\sigma^\gamma{}_\nu Y^\nu C_{\beta\gamma}{}^\alpha
+X^\mu\pa_\mu(\sigma^\alpha{}_\nu Y^\nu)-Y^\nu \pa_\nu(\sigma^\alpha{}_\mu X^\mu)
\nonumber\\
&=&
X^\mu Y^\nu\Big(\pa_\mu\sigma^\alpha{}_\nu-\pa_\nu\sigma^\alpha{}_\mu+C_{\beta\gamma}{}^\alpha\sigma^\beta{}_\mu \sigma^\gamma{}_\nu\Big)
+\sigma^\alpha{}_\mu \big[\un X,\un Y\big]^\mu
\eeqn
On the other hand, we know that $R^\sigma$ is vertical (recall eq. \eqref{Rsigisvert}), so 
\beqn
C_{\beta\gamma}{}^\alpha\sigma^\beta{}_\mu \sigma^\gamma{}_\nu=-\pa_\mu\sigma^\alpha{}_\nu+\pa_\nu\sigma^\alpha{}_\mu.
\eeqn
Similar to $\iota$, the $\sigma^\alpha{}_\mu$ are involved non-linearly in the relationship between the rotation coefficients in $H$ and those of $TM$ (which in a coordinate basis are zero). The connection implicit in the map $\sigma$ comes not from the $\sigma^\alpha{}_\mu$, but from the nature of the lift. We will explore the latter in the following subsection, which requires the precise notion of trivialization, and show how it is related to the connection coefficients introduced above.

\subsection{Trivializations}\label{sec:apptriv}

In this section we give some details on the trivialization of transitive Lie algebroids. As $A$ and its associated bundles are vector bundles over $M$, we need only to specify the nature of the transition functions employed on overlaps of coordinate patches in an atlas for $M$. While what discussed so far is valid for any transitive Lie algebroid, we will focus here on the specific case of Atiyah Lie algebroids.

The trivialization of $A$ involves specifying transition functions on overlaps of coordinate patches on $M$. Since globally $A=H\oplus V$, $H$ and $V$ are themselves vector bundles over $M$ and thus it is sensible to take transition functions that act on each bundle separately. On each coordinate patch $U_i\subset M$, we are to make a choice of basis on the fibre. If we choose a split basis for $A$, then the transition functions will be of the form given in eqs. (\ref{splitbasisV}-\ref{splitbasisH}), i.e.,\footnote{The convention adopted here is that the bundles, their bases and sections carry the patch subscript $U_i$ while the components of the various maps and sections have patch index $i$ only.}
\beq\label{transfnEAbasis}
 \un E_{\un A}^{U_i}=J_{ij}{}^{\un B}{}_{\un A}\un E_{\un B}^{U_j}, \qquad
 \un E_{\alpha}^{U_i}=J_{ij}{}^{\beta}{}_{\alpha}\un E_{\beta}^{U_j}.
\eeq
The transition functions on $TM$ are just the Jacobians associated with diffeomorphisms. That is
\beq
\un\pa^{U_i}_\mu={\cal J}_{ij}{}^\nu{}_\mu \un\pa^{U_j}_\nu,\qquad {\cal J}_{ij}{}^\nu{}_\mu\equiv\frac{\pa x^\nu_j}{\pa x_i^\mu},
\eeq
where $x_i^\mu$ are coordinates on the patch $U_i$, etc.
 The transition functions on any associated bundle are given by the corresponding group representation. For a vector bundle $E$ associated to a representation $R$ of the group $G$, choosing a basis for the fibre in each patch, we have
\beq\label{transfnEbasis}
\un e_a^{U_i}=R(g_{ij})^b{}_a\un e_b^{U_j},\qquad g_{ij}\in G.
\eeq
Correspondingly, the components of a section satisfy
\beq\label{transfnEcomp}
\psi^b_{j}=R(g_{ij})^b{}_a\psi^a_{i}.
\eeq
The latter comes simply from the tensorial property, $\un\psi_{U_i}=\un\psi_{U_j}$; eq. \eqref{transfnEcomp} is the familiar gauge transformation of a charged matter field. This applies equally well to the adjoint bundle, for which we write
\beq
\un t_A^{U_i}=t(g_{ij})^B{}_A\un t_B^{U_j}.
\eeq
For any associated bundle $E$, we have discussed the construction of $\hatd$ and the separation into horizontal and vertical parts, each of which is separately tensorial. For $\un\psi\in\Gamma(E)$, we had
$\hatd\un\psi(\un\mX)=\nabla^E_{\un\mX_H}\un\psi-v_E\circ\omega(\un\mX_V)(\un\psi)$ and given a trivialization we must have
\beq
(\nabla^E_{\un\mX_H}\un\psi)_{U_i}= (\nabla^E_{\un\mX_H}\un\psi)_{U_j}.
\eeq
Then, using \eqref{conncoeffsE}, we have
\beqn
\Big(\rho(\un\mX_H)(\psi^a_i)+A_i^a{}_b(\un\mX_H)\psi^b_i\Big)\un e_a^{U_i}
=\Big(\rho(\un\mX_H)(\psi^a_j)+A_j^a{}_b(\un\mX_H)\psi^b_j\Big)\un e_a^{U_j}.
\eeqn
Given \eqref{transfnEAbasis} and (\ref{transfnEbasis}-\ref{transfnEcomp}), it is straightforward to show that
\beq
A_i^b{}_c(\un E_\alpha^{U_i})=J_{ij}{}^\beta{}_\alpha\Big(
(R(g_{ij})^{-1})^b{}_d\rho(\un E_\beta^{U_j})(R(g_{ij})^d{}_c)
+(R(g_{ij})^{-1})^b{}_dA_j^d{}_b(\un E_\beta^{U_j})R(g_{ij})^b{}_c\Big),
\eeq
which is the expected transformation for a gauge connection.
We will discuss the tensorial nature of the vertical part below.
This result is true for any bundle associated to the Atiyah Lie algebroid $A=TP/G$, including the adjoint bundle $L$ itself. 

Although the above discussion constitutes a trivialization of the Lie algebroid, we have not made explicit the Ehresmann connection which $\sigma$ gives rise to. To do so, we note that locally (i.e., within any coordinate patch), the bundle $A$ can be thought of in terms of the bundle $L\oplus TM$. That is, on the patch $U_i$, we introduce a morphism \[\tau_i:A^{U_i}\to L^{U_i}\oplus TU_i,\]  with the property
\beq
\tau_i(\un\mX_H)=\mX_{i,H}^\alpha\tau_i{}^\mu{}_\alpha(\un\pa^{U_i}_\mu+b_{i}{}_\mu^A\un t^{U_i}_A),\qquad
\tau_i(\un\mX_V)=\mX_{i,V}^{\un A}\tau_i{}^A{}_{\un A}\un t^{U_i}_A.
\label{deftau}
\eeq
The map $\tau$ allows us to add sections of $TM$ and $L$ together, and gives an explicit expression for the split basis vectors. The $b_{i}{}_\mu^A$ are the components of the Ehresmann connection -- they parameterize the lift from $TM$ to $H$ given by the map $\sigma$. Indeed, as we will now show, the $b_{i}{}_\mu^A$ are equivalent to the connection coefficients discussed above. 
In fact, these properties follow because $\tau$ is a morphism. Expressions in the following discussion being local, they should be written with patch subscripts $U_i,i$. However, since here we work exclusively in the patch $U_i$,  we will for brevity drop these subscripts.

To begin, we first note that 
\beqn
\big[\tau(\un\mX_V),\tau(\un\mY_V)\big]_{L\oplus TM}&=&
\tau(\big[\un\mX_V,\un\mY_V\big]_A),
\eeqn
implies that
\beqn
\tau^A{}_{\un A}\tau^B{}_{\un B}f_{AB}{}^C
=C_{\un A\un B}{}^{\un C}\tau^C{}_{\un C},
\eeqn
which, making use of \eqref{rotstr}, can be written as
\beqn\label{VVtau}
(\tau\circ\iota)^A{}_{D}(\tau\circ\iota)^B{}_{E}f_{AB}{}^C
=(\tau\circ\iota)^C{}_{F}
f_{DE}{}^F,
\eeqn
where $(\tau\circ\iota)^A{}_B=\tau^A{}_{\un A}\iota^{\un A}{}_B$. 
Clearly, $\tau\circ\iota$ is a local endomorphism of $L$ and simply reflects the possibility of there being a change of basis for $L$ associated with the map $\tau$. Note that although $\tau^A{}_{\un A}$ has the same index structure as $\omega^A{}_{\un A}$, they are unrelated. Without loss of generality, we can take $\tau\circ\iota=Id_{L}$ (which since $\tau\circ\iota:L\to L\oplus TM$ means that we are taking this to be the trivial identification), at which point \eqref{VVtau} gives no further information, and the second of eq. \eqref{deftau} implies that $\tau:\iota(\un\mu)\mapsto (\un\mu,\un 0)$. 

Next we note that, again writing $\un\mY_V=\iota(\un\mu)$, and using \eqref{HVbracketnab},
\beqn
\big[\tau(\un\mX_H),\tau(\un\mY_V)\big]_{L\oplus TM}&=&\tau^\mu{}_\alpha\mX_H^\alpha\Big(\pa_\mu\mu^C+f_{AB}{}^Cb_\mu^A\mu^B\Big)\un t_C,
\\
\tau(\big[\un\mX_H,\un\mY_V\big]_A)&=&\tau\circ\iota(\nabla^L_{\un\mX_H}\un\mu)=\Big(\rho(\un\mX_H)(\mu^C)+\mX_H^\alpha A_\alpha{}^C{}_B\mu^B\Big)\un t_C,
\eeqn
So for $\tau$ to be a morphism,\footnote{Consequently, the map $\tau$ is determined by the local values of the $\iota$ and $\rho$ maps. However, we should emphasize that the $\rho$ and $\tau$ maps are different, as they have different image spaces, but the coefficients $\tau^\mu{}_\alpha$ and $\rho^\mu{}_\alpha$ are equal locally. The difference is precisely the Ehresmann connection coefficients $b^A_\mu$.} we are to set $\tau^\mu{}_\alpha=\rho^\mu{}_\alpha$, and then we obtain\footnote{Had we not assumed $\tau\circ\iota=Id_L$, then $A$ and $b$ would have been equal up to a shift proportional to $\pa_\mu(\tau\circ\iota)^A{}_B$. Note that locally we can model $L$ as $M\times\mg$ and consequently in a particular coordinate patch we can set $f_{AB}{}^C$ to be constant in an infinitesimal neighbourhood of a point. A similar statement can be made of $(t_A)^a{}_b$ for any associated bundle. Consequently, locally, it is possible to interpret the $\un t_A$ as elements of $\mg$ rather than $L\big|_{U_i}$.\label{footLoc}}
\beq\label{connisconn}
\rho^\mu{}_\alpha f_{AB}{}^Cb_\mu^A=A_\alpha{}^C{}_B.
\eeq
We see that the connection coefficients are equivalent to the Ehresmann connection components.

Finally, we note
\beqn\label{tautau}
\big[\tau(\un\mX_H),\tau(\un\mY_H)\big]_{L\oplus TM}&=&\big[\un X,\un Y\big]^\mu\un D_\mu+X^\mu Y^\nu F_{\mu\nu}^A\un t_A\\
\tau(\big[\un\mX_H,\un\mY_H\big]_A)&=&\tau(\sigma(\big[\un X,\un Y\big]))+\tau(R^\sigma(\un X,\un Y)),
\eeqn
where for brevity we have introduced $\un D_\mu\equiv \un\pa_\mu+b_\mu^A\un t_A$ and  have taken $\un\mX_H=\sigma(\un X)$, $\un\mY_H=\sigma(\un Y)$. In eq. \eqref{tautau}, we have denoted the components of the curvature of $b_\mu^A$ by 
\beq\label{curva}
F_{\mu\nu}^A=\pa_\mu b_\nu^A-\pa_\nu b_\mu^A+f_{BC}{}^Ab_\mu^B b_\nu^C,
\eeq
and the fact that $\tau$ is a morphism simply relates that to the other notions of curvature of the Lie algebroid, in agreement with \eqref{connisconn}.

So far, we have demonstrated that the connection on the Lie algebroid appears within the geometric structure in several equivalent ways, consistent with our previous demonstration that the curvature also  appears in several equivalent ways as well. The local trivialization has just uncovered one more way to see this, based on the explicit extraction of the Ehresmann connection. There is one remaining aspect that we did not directly address in the above discussion, which is the vertical part of $E$-valued extended forms. Consider the one form $\hat d\un\psi\in\Gamma(A^*\times E)$. Its vertical part is tensorial, and given by (see \eqref{brstsecE})
\beq\label{vertparthatdpsi}
s\secE(\un\mX)=\hatd\secE(\un\mX_V)=-v_E\circ\omega(\un\mX_V)(\secE)=-\ghost_E(\secE)(\un\mX_V).
\eeq
In this expression, $\omega$ is an $L$-valued 1-form. In a local trivialization, restoring momentarily the patch subscripts, we can then write
\beq
\omega_{U_i}=\omega_i^A{}_{\un A} E_{U_i}^{\un A}\otimes \un t_A^{U_i}.
\eeq
To understand this, we need to understand the local forms $E_{U_i}^{\un A}$, the dual basis in $A^*|_{U_i}$. Given the map $\tau_i$ introduced above, we can deduce their structure from eq. \eqref{dualformorth}. The result is
\beq\label{basE}
\tau^*_i(E_{U_i}^\alpha)=(\tau_i^{-1}){}^\alpha{}_\mu dx_i^\mu,\qquad
\tau^*_i(E_{U_i}^{\un A})=(\tau_i^{-1})^{\un A}{}_A(t_{U_i}^A-b_{i}{}_\mu^A dx_i^\mu),
\eeq
where $\tau^*:A^*\to L^*\oplus T^*M$ is the map dual to $\tau$ (preserving the pairing \eqref{dualformorth}) and $t_{U_i}^A$ is a basis form for $L^*_{U_i}$. So we see that locally, the form $\omega_{U_i}$ contains two things: the coefficient $\omega_i^A{}_{\un A}$, and the connection coefficients $b_{i}{}_\mu^A$. The latter is responsible for the fact that $\Omega$ coincides with the curvature $R^\sigma$. Again, in what follows we will drop the patch subscripts for brevity. We interpret the coefficients $\omega{}^A{}_{\un A}$ as the components of the ghost field, which is in line with \eqref{vertparthatdpsi}, encoding the BRST transformation of the field $\secE$. Locally, this appears as
\beq
s\secE(\un E_{\un A})=-\omega^A{}_{\un A}v_E(\un t_A)(\un\psi)=-(t_A)^a{}_b\omega^A{}_{\un A}\psi^b\un e_a,
\eeq
which we can interpret as a form equation\footnote{In standard physics notation, one dispenses with the basis vectors and writes $s\psi^a=-c^a{}_b\psi^b$. We should note though that this is an interpretation: our $s$ actually acts linearly, so $s\secE=s(\psi^a\un e_a)=\psi^a s(\un e_a)=-\psi^a c^b{}_a\un e_b$. }
\beq\label{cpsi}
s\secE=s(\psi^a\un e_a)=-\ghost^a{}_b\psi^b\un e_a=-\ghost_E(\secE),
\eeq
with the vertical form gauge ghost field $c^a{}_b$  given by
\beq\label{ghost}
\ghost^a{}_b\equiv \ghost_{\un A}{}^a{}_bE^{\un A}=(t_A)^a{}_b \omega^A{}_{\un A} E^{\un A}.
\eeq
Eq. \eqref{cpsi} is the usual form of a BRST transformation in terms of the ghost field, with its usual Grassmann nature replaced by it being a vertical form in $A^*$.
The same interpretation of the Grassmann nature of the ghost was firstly proposed in \cite{Neeman:1979cvl, ThierryMieg:1979xe, ThierryMieg:1979kh, Baulieu:1981sb, ThierryMieg:1982un, ThierryMieg:1987um} in the context of principal bundles, see Section \ref{sec:physicsTPmodG}. We have here realized it in the framework of Lie algebroids.

%This aligns with the idea of Thierry-Mieg and collaborators \cite{Neeman:1979cvl, ThierryMieg:1979xe, ThierryMieg:1979kh, Baulieu:1981sb, ThierryMieg:1982un, ThierryMieg:1987um} discussed in Section \ref{sec:physicsTPmodG} but arises as a matter of course in the framework of Lie algebroids. 

Eq. \eqref{ghost} can be equivalently recast as an expression for an extended one form valued in $End(E)$:
\beq
v_E\circ\omega=(t_A)^a{}_b \omega^A{}_{\un A}E^{\un A}\otimes \un e_a\otimes f^b=c_{\un A}{}^a{}_b E^{\un A}\otimes \un e_a\otimes f^b=c^a{}_b \un e_a\otimes f^b=c_E.
\eeq
To recap, we have identified locally the gauge connection and have shown that it is equivalently described as an Ehresmann connection. We then proved that the gauge ghost field is encoded in the components of the connection reform. 
 We will expand more on the physics of this in subsection \ref{rec}.

\subsection{BRST transformation of gauge fields and ghosts}\label{sec:BRSTgaugeghost}

In the previous section we have related the local component field $\omega_i^A{}_{\un A}$ with the gauge ghost, and noted that the BRST transformation of any section of a bundle $E$ properly involves it. In this section, we will show that the BRST transformations of the ghost and gauge fields are implicitly built into the formalism as well. 
The BRST transformation of the ghost field itself comes directly from the result that $\Omega$ is horizontal. That is, from \eqref{defcurvatureform}, the vertical part of $\hatd\omega$ is simply $-\frac12[\omega,\omega]$. Applying the morphism $v_E$ to \eqref{defcurvatureform} and using the linearity of $s$ we obtain that the BRST transformation of $c_E$ is given by
\beq
s c_E=s(v_E\circ \omega)=v_E\circ (s\omega)=v_E(-\tfrac12\big[\omega,\omega])=-\tfrac12\big[v_E\circ \omega,v_E\circ \omega]=-\tfrac12\big[c_E,c_E].
\eeq
For its $End(E)$ components this equation reads
\beq\label{sofg}
s\ghost^a{}_b=-\tfrac12\big[\ghost,\ghost]^a{}_b,
\eeq
which is the well-known BRST transformation of the gauge ghost in gauge theories  \cite{Becchi:1975nq, Tyutin:1975qk}.

The BRST transformation of the gauge field follows from the nilpotency of $\hatd$ and thus is also automatic. 
From \eqref{dpsi} for $n=1$, consider the case
\beq
\hat d \secE_1 (\un\mX_V,\un\mY_H) =
(\nabla^{A^*\times E}\secE_1)(\un\mX_V,\un\mY_H)
+s\secE_1(\un\mX_V,\un\mY_H).
\eeq
If we take $\secE_1=\hat d\secE_0$ then this is just automatically zero by nilpotency,
that is,
\beq\label{TNT}
(s\hat d\secE_0)(\un\mX_V,\un\mY_H)=-(\nabla^{A^*\times E}\hat d\secE_0)(\un\mX_V,\un\mY_H).
\eeq

We now compute the two sides separately. For the left-hand side, we have
\beqn
(s\hatd\secE_0)(\un \mX_V,\un\mY_H)
=-s(\hatd\secE_0(\un\mY_H))(\un\mX_V)
=-s(\nabla^E_{\un\mY_H} \secE_0)(\un\mX_V).\label{sdpsi}
\eeqn
For the right-hand side we use \eqref{nabAETot} and compute
\beqn
(\nabla^{A^*\times E}\hat d\secE_0)(\un\mX_V,\un\mY_H)=-(\nabla^{A^*\times E}_{\un\mY_H}\hat d\secE_0)(\un \mX_V)=
-\nabla^E_{\un\mY_H}(\hat d\secE_0(\un\mX_V))
+(\hat d\secE_0)(\big[\un\mY_H,\un\mX_V\big]_A).
\eeqn
Since the commutator in the last term is vertical, only the vertical part of $\hat d\secE_0$ contributes
\beq
(\nabla^{A^*\times E}\hat d\secE_0)(\un\mX_V,\un\mY_H)=\nabla^E_{\un\mY_H}(c_E(\secE_0)(\un\mX_V))
-c_E(\secE_0)(\big[\un\mY_H,\un\mX_V\big]_A)=(\nabla_{\un\mY_H}^{A^*\times E}c_E(\secE_0))(\un\mX_V).
\eeq
Notice that, using the nilpotency of $\hat d$ for two horizontal sections of $A$, we can replace $\un\mX_V$ with the full $\un\mX$ in this expression. Since we can do the same also in \eqref{sdpsi}, eq. \eqref{TNT} becomes
\beq
s(\nabla^E_{\un\mY_H} \secE_0)(\un\mX)=(\nabla_{\un\mY_H}^{A^*\times E}c_E(\secE_0))(\un\mX).
\eeq
Specializing to $\secE_0\to\un e_a$ and $\un\mY_H\to\un E_\alpha$, and writing the corresponding form equation in $A^*$, we obtain
\beq
s\nabla^E_{\un E_\alpha}\un e_a=
\nabla^{A^*\times E}_{\un E_\alpha}\ghost_E(\un e_a).
\eeq
Given eq. \eqref{conncoeffsE}, we have
\beq\label{BRSTA}
s(A_{\alpha}{}^b{}_a\un e_b)
=
\nabla^{A^*\times E}_{\un E_\alpha}\ghost^b{}_a\un e_b,
\eeq
where we used \eqref{cpsi} and the notation $A_\alpha{}^b{}_a=A^b{}_a(\un E_\alpha)$.
This is the familiar result that the BRST transformation of a gauge field is given by the covariant derivative of the gauge ghost \cite{Becchi:1975nq, Tyutin:1975qk}. We remark again that this equation and the one establishing the BRST transformation of the ghost are an automatic consequence of our geometric framework, where we have not imposed any external condition, in contrast with the original accounts on geometrization of BRST \cite{Neeman:1979cvl, ThierryMieg:1979xe, ThierryMieg:1979kh, Baulieu:1981sb, ThierryMieg:1982un, ThierryMieg:1987um}, where these equations follow from restricting to right invariant quantities on a principal bundle.
% the externally imposed requirement of the extended curvature being horizontal. 
This is clearly an advantage of the algebroid geometry showing how it already contains all the useful structure for physical applications, and nothing more. 

\subsection{Back to physics}\label{rec}

In this section, we have translated much of the important structure of field theories based on Atiyah Lie algebroids into a local index notation. As mentioned earlier in Sec. \ref{sec:physicsTPmodG}, one can then express any gauge-matter theory in its terms. We have seen that BRST is simply built into the formalism geometrically, and all of the familiar properties are obtained directly. We should note that we have made no mention so far of a classical action, or a gauge fixing prescription, which are often given as important ingredients for BRST constructions \cite{Becchi:1975nq, Tyutin:1975qk, Gribov:1977wm, Bonora:1982ve,  Gregoire:1992tf, Grigoriev:2006tt, Upadhyay:2010ww}. Here, we have a very different attitude: the BRST structure, that is the gauge covariance, is simply built into the geometry of the theory, and any sensible choice of action must be invariantly defined. In local terms, this translates into the action being gauge invariant. We should also note that the construction is inherently diffeomorphism invariant as well, which again is possible because we have not specified auxiliary structure such as a metric. Of course, traditionally in Yang-Mills theories, in order to write an action one must supply additional geometric details, usually a choice of metric. The diffeomorphism invariance of the resulting action is only broken in the sense that we usually think of the metric as fixed. In this sense then, diffeomorphism invariance is ``spontaneously broken'', and it is important that we have an off-shell notion of the full geometric structure.

Before moving on to gravitational theories, we would like to offer here a summary of our results and offer some guidance to their use. Thanks to the index notation just introduced we can now make full contact with typical discussions on gauge theories in physics. The novelty is that, on top of seeing how usual gauge theory quantities are related to  various geometric objects on Atiyah Lie algebroids,\footnote{As stated earlier, the formulation of gauge theories on Atiyah Lie algebroids is also addressed in \cite{Lazzarini_2012, Fournel:2012uv, Jordan:2014uza, Carow-Watamura:2016lob, Kotov:2016lpx, Attard:2019pvw}.} we can do the same also for BRST data. 

The basic ingredient of a gauge theory is its gauge Lie group $G$. A gauge group is accounted for on a principal fibre bundle $P$, as described in Section \ref{sec:princbdl}. The second key ingredient is a gauge connection. As we reviewed, although one can define it in $P$, the natural arena for gauge connections is in fact the Atiyah Lie algebroid $TP/G$. There, the redundant right group action is modded out giving rise to a bundle over $M$.  In physical applications,  a gauge connection is usually thought of simply as a one form on $M$ transforming (non-linearly) in the adjoint representation of the Lie algebra $\mg$ associated to the gauge Lie group $G$. This quantity can be locally identified in our construction with $b^A_\mu$, which enters the discussion in eq. \eqref{deftau}, (see footnote \ref{footLoc}). The Atiyah Lie algebroid has by construction a Lie algebra adjoint action at each point in $M$, related to the bundle $L_G$ in the short exact sequence defining $TP/G$. The gauge connection is truly an Ehresmann connection needed to single out a horizontal distribution on the Lie algebroid. As shown in \eqref{connisconn}, the gauge connection is equivalent to the connection appearing in the gauge covariant derivative. In addition to the  bundle $L_G$, on which the  connection coefficients $A_\alpha{}^B{}_A$ are defined via \eqref{AinL}, we also find the induced connection coefficients on any associated bundle. We see thus that the gauge connection manifests itself in various places: as an Ehresmann connection $\sigma$, locally as a building block of the basis in $H$ and as the connection in the gauge covariant derivative on associated bundles. The curvature of a gauge connection is its field strength, which we typically write as in \eqref{curva}. It is the gauge covariant quantity used to build invariant actions. The geometric picture developed here relates it to the 
curvature of the map $\sigma$. This is furthermore encoded in the curvature two form of the connection reform $\omega$. A therefore equivalent way to express the gauge curvature is as the non-integrability of the distribution $H$. Since there are various maps and bundles involved in our geometry, there are various notions of curvature. Part of our agenda was to show that they are all intertwined, as different realizations of the gauge curvature, the geometric origin being in the curvature of the Ehresmann connection, controlling the non-integrability of $H$ in $TP/G$.

In the last paragraph, we have discussed a gauge theory without matter -- the sections $\un\mX$ of $A$ and $\un\mu$ of $L_G$ are not fields, but just useful tools employed in the description of the Lie algebroid and its connection. Physically, charged matter fields are quantities transforming linearly in some representation $R$ of the gauge group $G$. Geometrically, these fields are just the components $\psi^a$ of an associated bundle $E$ via the representation $R$, as they are in the usual formalism of principal bundles, see e.g. \cite{RevModPhys.52.175}. There is then an induced gauge covariant derivative $\nabla^E$ acting on the associated bundle, which we have found as the horizontal part of the exterior derivative $\hat d$ defining the complex $\Omega^\bullet(A,E)$. The split into vertical and horizontal parts of the exterior derivative on extended forms in $\Omega^\bullet(A,E)$ was an important achievement, for the vertical part, discussed shortly, contains relevant information as well. This completes the list of ingredients necessary to construct classical gauge theories, and all the necessary data is available geometrically in a relatively simple way on Atiyah Lie algebroids and associated bundles. 

 The classical geometry developed in this paper, as in \cite{Neeman:1979cvl, ThierryMieg:1979xe, ThierryMieg:1979kh, Baulieu:1981sb, ThierryMieg:1982un, ThierryMieg:1987um}, also controls the quantized theory, because the BRST formalism is naturally included. In the familiar formalism, as reviewed for instance  in \cite{Nemeschansky:1987xb, Henneaux:1989jq, Henneaux:1992ig, Becchi:1996yh, VanHolten:2001nj, Fuster:2005eg}, the Fadeev-Popov ghosts appear through a gauge-fixing procedure in the quantization of gauge theories, and the BRST symmetry is the symmetry of a gauge-fixed Lagrangian.  The details of the BRST procedure are usually thought of as dependent on the choice of the latter, but since the origin of the BRST operator is geometrical, one should be able to deal with it off-shell. This is the direction we followed here, where everything is independent of on-shell dynamics and relies only on geometric structure. 
Different Lie algebroids, with a given structure group and base $M$, are distinguished precisely by different connections, so integrating over connections can be regarded as the same thing as summing over Lie algebroid geometries.

One of the important properties of the BRST ghost is that it is Grassmann-valued. Here, the Grassmann algebra is geometrically realized as an exterior algebra. Although this statement resembles that of \cite{Neeman:1979cvl, ThierryMieg:1979xe, ThierryMieg:1979kh, Baulieu:1981sb, ThierryMieg:1982un, ThierryMieg:1987um}, the details are somewhat different and in fact globally well-defined, because of our implementation of this idea on Atiyah Lie algebroids. The Fadeev-Popov ghost has the geometric and gauge properties compatible with $c^a{}_b$ introduced in \eqref{ghost}. The Grassmann nature is accounted for by the fact that it is a vertical one form in $A^*$. 
It is often said heuristically that the presence of the Fadeev-Popov ghosts is responsible for the correct number of degrees of freedom contributing to the quantum path integral; apparently then, the connection on an Atiyah Lie algebroid has built-in the correct degrees of freedom. All of these are packaged together into the connection reform; on the one hand, its components $\omega^A{}_{\un A}$ in a local basis are related to the ghost \eqref{ghost}, while on the other hand, it locally carries information on the gauge connection via the basis itself in $A^*$, eq. \eqref{basE}. 
It is important that its curvature two-form is yet another manifestation of the gauge curvature, horizontal by construction.

So given a gauge group, we know how to geometrically establish in $TP/G$ a gauge field and gauge ghosts. The 
last ingredient we need is the BRST operator $s$. In the usual gauge theory quantization, this operator generates the BRST symmetry \cite{Becchi:1975nq, Tyutin:1975qk}. Here, we have shown that it is encoded in the vertical part of the exterior derivative $\hat d$ on Lie algebroids or associated bundles, eq. \eqref{brstsecE}. The globally well-defined distinction between vertical and horizontal parts of $\hat d$ therefore distinguishes between BRST and covariant gauge derivative, respectively. 
For instance, the famous Darboux-Maurer-Cartan-Ehresmann structure equation of the principal fibre bundle \cite{zbMATH02632856, spivak1979comprehensive}, stating that the curvature two form is horizontal, and giving rise to the Russian formula \cite{Neeman:1979cvl, ThierryMieg:1979xe, ThierryMieg:1979kh, Baulieu:1981sb, ThierryMieg:1982un, ThierryMieg:1987um}, is here just a straightforward geometrical consequence of Atiyah Lie algebroids, rather than a consequence of imposing $G$-equivariance of the connection.
%a constraint to be imposed. 
The fact that $s$ can be interpreted as the BRST operator has been clearly established by observing that the components of the matter fields $\psi^a$, ghosts $c^a{}_b$ and gauge connections $A_\alpha{}^a{}_b$ transform under the BRST symmetry as is well-known in physics, eqs. \eqref{cpsi}, \eqref{sofg} and \eqref{BRSTA}. Again, the placement of $s$ as the vertical part of an extended notion of exterior derivative is a familiar part of the usual  
geometric construction of BRST 
\cite{Neeman:1979cvl, ThierryMieg:1979xe, ThierryMieg:1979kh, Baulieu:1981sb, ThierryMieg:1982un, ThierryMieg:1987um}, whereby one formally adds commuting and anti-commuting quantities together. 
On Atiyah Lie algebroids, this construction has a well-defined geometric origin, as we have thoroughly discussed.

It is well-known that the cohomology of the extended exterior derivative controls the structure of anomalies as well as many other features of gauge theories \cite{Bonora:1982ve, Zumino:1984ws, Bandelloni:1986wz, Henneaux:1989rq, Brandt:1989gv, Henneaux:1995ex, Dragon:1996md, DuboisViolette:1991is, Barnich:1994db, Barnich:1994mt, Barnich:1995ap, Brandt:1996mh, Barnich:2000zw}. Here, the same will be true for $\hatd$, and we have noted the structure of its complex throughout the paper. We will not explore this further in this paper, leaving it for future works. In the next section we will discuss gravitational theories from the point of view of Lie algebroids, and show that it is another application of the general construction we developed above.

\section{Gravitational Theories}\label{sec:grav}

Given that the Lie algebroid construction is intrinsically defined with respect to both diffeomorphisms and group invariance, it is natural to consider gauge theories that have a gravitational interpretation. 
In this section then, we discuss the features of Lie algebroids and associated bundles that are relevant to gravitational theories. Necessarily, we will arrive at such theories in a ``first order'' form; that is, the principal bundle of interest is the frame bundle over a $d$-dimensional manifold $M$, and we will introduce a Lie algebroid with structure group given by a subgroup of $GL(d,\mathbb{R})$ along with a connection on the Lie algebroid. In addition, we must introduce an associated bundle whose sections are given by a solder form.

At the level of the principal frame bundle, restricting to a subgroup $G\subset GL(d,\mathbb{R})$ means that we are considering a particular $G$-structure, which we denote by $F_G$. Often, we can associate a choice of $G$-structure to a requirement that a geometric structure remain invariant \cite{chern1966, molino1972, Godina_2003, kobayashi1995transformation}. The usual case of interest is to take $G$ to be the local Lorentz group, $G=SO(1,d-1)$, which is associated with the requirement of preserving a metric (frames remaining orthonormal). Such an interpretation is an auxiliary notion, and we will not make direct use of it here. So for now we simply assume that we have some subgroup $G\subset GL(d,\mathbb{R})$ of dimension $\dim G$.\footnote{Here we are supposing that $G$ at least contains $SO(1,d-1)$. There is no obstruction to selecting smaller subgroups. Fashionable examples would include those with non-relativistic (Galilean) or ultra-relativistic (Carrollian) symmetry \cite{Duval:1984cj, Figueroa-OFarrill:2018ilb, Figueroa-OFarrill:2020gpr}.} As discussed earlier in the paper, Sec. \ref{sec:atiyah}, we can then construct an Atiyah Lie algebroid from $F_G$ by quotienting $TF_G$ by the right action of $G$. We will refer to this Lie algebroid as $A_G=TF_G/G$ and the adjoint bundle as $L_G$, 
\beq\label{splitshortExactSeq}
\begin{tikzcd}
&
0
\arrow{r} 
& 
L_G
\arrow{r}{\iota}
& 
A_G
\arrow{r}{\rho} 
& 
TM
\arrow{r}
&
0
\end{tikzcd}.
\eeq
Having introduced a connection for $A_G$, given the above results, we can anticipate that there will be an Ehresmann connection with components $b_\mu^A$, and a $G$-ghost field determined by $\omega^A{}_{\un A}$. 

We will now suppose that there is a $d$-dimensional representation $R_G$ which gives rise to an associated vector bundle ${\cal E}=F_G\times_{R_G} \mathbb{V}$ of rank $d$, where $\mathbb{V}$ is the vector space of the representation $R_G$. In general, the dual bundle ${\cal E}^*$ is inequivalent. A \hlt{solder form} gives an isomorphism $\solder:TM\to {\cal E}$. This defines the familiar notion of a moving frame. In simple terms, suppose we take a coordinate basis $\{\un\pa_\mu\}$ of sections of $T\cM$, and a basis of sections $\{\un e_a\}$ for ${\cal E}$. Then, we may write 
\beq\label{solderaction}
\solder(\un\pa_\mu)=\solder^a_{\mu}\,\un e_a.
\eeq
We think of the coefficients as giving a 1-form on $M$ transforming in $R_G^*$,
\beq
\solder^a=\solder^a_\mu dx^\mu,
\eeq
That is, we can equivalently regard $\solder\in\Gamma(T^*M\times{\cal E} )$, and write
\beq
\solder =  \solder^a\un e_a.
\eeq
As such, $\theta$ is a section of a bundle of a type that we have not considered before. However, given that ${\cal E}$ is an  associated  bundle to the Lie algebroid $A_G$, we can make  use of the anchor map $\rho$ for $A_G$ to introduce the closely related map
\beq
\hat\solder=\theta\circ\rho,\qquad\qquad \hat\solder:A\to {\cal E}.
\eeq
We can regard $\hat\solder\in\Gamma(A^*\times {\cal E})$, which is a section of a bundle of the type that we have considered throughout the paper. We will refer to $\hat\solder$ generally as the solder form. We note that because of its definition, it is automatically horizontal. This is consistent with both our results and the literature. As we have established, given that $\hat\solder$ is a section of an associated bundle, we know how it transforms under the action of the  BRST operator. In some accounts of BRST in the case of $G=SO(1,d-1)$ (e.g. \cite{Baulieu:1984pf, Baulieu:1983tg, Baulieu:1985md}), it is often assumed that the coefficients $\solder^a$ has no vertical part, which then fixes its BRST transformation, thanks to the requirement of the extended torsion being horizontal. Here once again, this is a built-in feature of the Lie algebroid geometric structure, similarly to what happens with the Russian formula and the BRST transformation of the gauge field and ghosts, eqs. \eqref{sofg} and \eqref{BRSTA}.

In gravitational theories, we would interpret $\hat\solder$ as a ``matter'' field in that it is a section of an associated bundle, rather than being related to the connection on $A_G$. The latter, for $G=SO(1,d-1)$, contains the Lorentz ghost and the usual Lorentz  spin  connection, as we will shortly unveil. Together, this connection and solder form are regarded as equivalent to a more traditional form of a gravitational theory, which of course usually involves a metric on $TM$. In a given theory, this equivalence might come about through assumption (by introducing constraints) or classically as a result of equations of motion. In metric theories, it is well-known that there is a unique connection on $TM$ (the Levi-Civita connection \cite{LC}), which is both metric-compatible and torsion-free. In the present context, the metric compatibility can be interpreted as a consequence of choosing $G=SO(1,d-1)$, while the torsion-free condition comes about if it is possible for the solder form to be covariantly constant. 

Given the solder form, we can also consider the dual bundle ${\cal E}^*$. Here, we mean the vector bundle associated with $A_G$ through the dual representation $R_G^*$, and we introduce  $\dsolder:TM\to {\cal E}^*$, as well as $\hat\dsolder=\dsolder\circ\rho:A\to {\cal E}^*$, which we refer to as the \hlt{Schouten form}.\footnote{We use this name because the usual notion of the Schouten tensor can be incorporated into the Lie algebroid language as a section of $A^*\times {\cal E}^*$. Indeed in this context, $\hat\solder$ and $\hat\dsolder$ are not related, the former giving a choice of frame and the latter, in fact, related to curvature. We will report more on this in subsequent papers.} Depending on the choice of $G$, ${\cal E}$ and ${\cal E}^*$ may or may not be equivalent.  Nevertheless, since ${\cal E}$ and ${\cal E}^*$ are dual bundles, we regard ${\cal E}^*:{\cal E}\to C^\infty(M)$. Following the general notation introduced earlier (Sec. \ref{Sec4} and Appendix \ref{sec:notation}), we will denote by $\{ f^a\}$ a local basis for ${\cal E}^*$ which is dual to the basis $\{\un e_a\}$ for ${\cal E}$,
$f^a(\un e_b)=\delta_b{}^a$.

The structures that we have discussed are summarized in the following figure.
\beq\label{splitshortExactSeq}
\begin{tikzcd}
&
&
&
{\cal E}
&
\\
0
\arrow{r} 
& 
L_G
\arrow{r}{\iota} 
\arrow[bend left]{l} 
& 
A_G
\arrow{r}{\rho} 
\arrow{ur}{\hat\solder}
\arrow[swap]{dr}{\hat\dsolder}
\arrow[bend left]{l}{\omega}
& 
TM
\arrow[swap]{u}{\solder}
\arrow{d}{\dsolder}
\arrow{r} 
\arrow[bend left,swap]{l}{\sigma}
&
0
\arrow[bend left]{l} 
\\
&&&{\cal E}^*&
\end{tikzcd}.
\eeq
The central part of this figure is identical to any Atiyah Lie algebroid, whereas the associated bundles are specific to pure gravitational theories. In the following section, we discuss some features of the solder form and the action of the  exterior derivative  on it.

\subsection{Solder and torsion}\label{sec:solder}

Given that we have described the solder form as a section of the associated bundle ${\cal E}$, it is just an application of the general results of the previous sections to describe how $\hatd$ acts here. If we regard $\hat\solder$ as a section of $A^*\times {\cal E}$, and recall that $\hat\solder(\un\mX_V)=0$, then we immediately have
\beqn
(\hatd\hat\solder)(\un\mX,\un\mY)
&=&\phi_{{\cal E}}(\un\mX)(\hat\solder(\un\mY))-\phi_{{\cal E}}(\un\mY)(\hat\solder(\un\mX))
-\hat\solder(\left[\un\mX,\un\mY\right]_A)
\label{hatdtheta}
\\
&=&
\nabla^{{\cal E}}_{\un\mX_H}\hat\solder(\un\mY)
-\nabla^{{\cal E}}_{\un\mY_H}\hat\solder(\un\mX)
-\hat\solder(\left[\un\mX,\un\mY\right]_A)
-v_{{\cal E}}\circ\omega(\un\mX_V)(\hat\solder(\un\mY_H))
+v_{{\cal E}}\circ\omega(\un\mY_V)(\hat\solder(\un\mX_H)),
\eeqn 
where $v_{{\cal E}}\circ\omega\in\Gamma(A^*\times End({\cal E}))$ and  $\hat\solder(\un\mY_H)\in\Gamma({\cal E})$. 
The last two terms can then be rewritten
\beq
v_{{\cal E}}\circ\omega(\un\mX_V)(\hat\solder(\un\mY_H))
-v_{{\cal E}}\circ\omega(\un\mY_V)(\hat\solder(\un\mX_H))
=(v_{{\cal E}}\circ\omega\wedge\hat\solder)(\un\mX,\un\mY),
\eeq
where the wedge product is taken in $A^*$.
The torsion is then defined as usual as the covariant derivative
\beqn
\hat T(\un\mX,\un\mY)\label{deftorsion}
&=&(\hatd\hat\solder+v_{{\cal E}}\circ\omega\wedge\hat\solder)(\un\mX,\un\mY)\\
&=&
\nabla^{{\cal E}}_{\un\mX_H}\hat\solder(\un\mY)
-\nabla^{{\cal E}}_{\un\mY_H}\hat\solder(\un\mX)
-\hat\solder(\left[\un\mX,\un\mY\right]_A)
\\
&=&(\nabla^{A^*\times{\cal E}}\hat\solder)(\un\mX,\un\mY),
\eeqn
the last equality being an immediate consequence of the results in Appendix \ref{app:LAmaps}.
As we mentioned earlier, since $\hat\solder$ is a section of an associated bundle (and not of $A$ itself), it has no notion of vertical and horizontal, and simply annihilates the vertical part of a section of $A$. Consequently, we immediately obtain
\beqn
\hat T(\un\mX,\un\mY)\label{hortorsion}
&=&
\nabla^{{\cal E}}_{\un\mX_H}\hat\solder(\un\mY_H)
-\nabla^{{\cal E}}_{\un\mY_H}\hat\solder(\un\mX_H)
-\hat\solder(\big[\un\mX_H,\un\mY_H\big]_A),
\eeqn
and so the torsion is manifestly and automatically horizontal. Notice that in the last term, since $\hat\solder$ is horizontal, the curvature does not enter.
 In a local trivialization, eq. \eqref{hortorsion} evaluates to 
\beq
\hat T(\un\mX,\un\mY)=
\mX_H^\alpha\mY_H^\beta \rho^\mu{}_\alpha\rho^\nu{}_\beta\Big(\pa_\mu\solder^a_\nu+A_\mu{}^a{}_b\solder^b_\nu-\pa_\nu\solder^a_\mu
-A_\nu{}^a{}_b\solder^b_\mu
\Big)\un e_a.
\eeq
In the case of $G$ being the Lorentz group, $A_\mu{}^a{}_b$, with $A_\alpha{}^a{}_b=\rho^\mu{}_\alpha A_\mu{}^a{}_b$, are the components of  the Lorentz spin connection.
In previous accounts of the topic \cite{Baulieu:1984pf, Baulieu:1983tg,Baulieu:1985md}, the %frame field 
solder form has been endowed with a Grassmann partner, and the torsion then required to be horizontal, to obtain the BRST transformation of $\hat\theta$. %frame.
Here we see that geometrically, there is no sensible notion of the former, and the latter is automatically satisfied, 
further indicating how the Atiyah Lie algebroid formalism is the correct framework to address also gravitational theories.

As well, the BRST transformation is simply encoded in $\hatd\hat\solder$, which we deduce by considering the horizontal/vertical parts as follows
\beqn
(\hatd\hat\solder)(\un\mX_H,\un\mY_H)
&=&
\nabla^{{\cal E}}_{\un\mX_H}\hat\solder(\un\mY_H)
-\nabla^{{\cal E}}_{\un\mY_H}\hat\solder(\un\mX_H)
-\hat\solder(\big[\un\mX_H,\un\mY_H\big]_A)
=\hat T(\un\mX_H,\un\mY_H),\\
(\hatd\hat\solder)(\un\mX_V,\un\mY_V)
&=&
\un 0,
\\
(\hatd\hat\solder)(\un\mX_V,\un\mY_H)
&=&
-v_{{\cal E}}\circ\omega(\un\mX_V)(\hat\solder(\un\mY_H))
=-c_{{\cal E}}(\hat\solder(\un\mY_H))(\un\mX_V).
\eeqn
Since $\hat\solder$ is horizontal, the latter equation corresponds to the BRST transformation $s\hat\solder$; in a local basis, it evaluates to 
\beqn
s\hat\solder(\un E_{\un A},\un E_\alpha)
=-\ghost_{{\cal E}}(\hat\solder(\un E_\alpha))(\un E_{\un A})=-\rho^\mu{}_\alpha \ghost_{\un A}{}^a{}_b\solder^b_\mu\un e_a,
\eeqn
or, as a form equation,
\beqn
s\hat\solder=-\ghost^a{}_{b}\wedge\hat\solder^b\un e_a.
\eeqn
Here, $\hat\solder^b=\rho^\mu{}_\alpha \theta^b_\mu E^\alpha$, and $\ghost^a{}_{b}$ is the usual Lorentz ghost if $G$ is the Lorentz group.
Similarly, we can apply all of the above to the Schouten form $\hat\dsolder$. In that case, the analogue of the torsion is the Cotton form.

\subsection{Diffeomorphisms and diffeomorphism ghosts}\label{sec:diffghosts}

In BRST discussions of gravitational theories, the diffeomorphism ghost plays a prominent role. This is introduced to account for the invariance under diffeomorphisms of such theories. However, it should be noted that at least classically, it is possible to decouple the diffeomorphism ghost by redefinitions, as was explained first in \cite{Baulieu:1984pf, Baulieu:1983tg, Baulieu:1985md}. 

Throughout the paper, we made no reference to a diffeomorphism ghost; indeed, the ghosts appeared as vertical forms encoded in the $\omega$ map, and so are associated with the structure group of the Lie algebroid. That is, we saw that only gauge ghosts arise in the formalism. The construction is intrinsically well-defined, and thus independent of a choice of local coordinates. One might imagine that if diffeomorphisms were somehow promoted to be part of the structure group, then the diffeomorphism ghosts would appear.\footnote{Given our geometric formalism, we regard the BRST operator to act on diffeomorphism invariant quantities, as opposed to covariant quantities, such as the local components of a form in a given basis.} 

We do not advocate this picture here. Instead, we note that the result quoted above of Refs. \cite{Baulieu:1984pf, Baulieu:1983tg, Baulieu:1985md} is consistent with the fact that we do not have diffeomorphism ghosts. As we now describe, there is a natural place for the diffeomorphism ghosts to appear within the Lie algebroid formalism. Recall that this ghost, in the traditional description, is thought of as a Grassmann vector, $\zeta^\mu$. Here that description would carry over to a vector-valued vertical form, a tensor with index structure $\zeta^\mu_{\un A}$. Whereas, the gauge ghosts appear in $\omega: A\to L$, one might expect that diffeomorphism ghosts should analogously appear in $\rho: A\to TM$. Indeed, as we described in Section \ref{Sec4}, the $\rho$ map does have generally the index structure $\rho^\mu{}_M$, and thus in the language of a split basis, there are the components $\rho^\mu{}_{\un A}$. However, those components were set globally to zero in any split basis, as a result of the $\un E_{\un A}$ spanning the vertical sub-bundle of $A$, the kernel of $\rho$. We thus arrive at a simple picture: given a connection on the Lie algebroid, it was possible to define a split basis; such a choice can be thought of as the choice that absorbs the diffeomorphism ghost into a redefinition. It is important that the curvature of the map $\rho$ vanishes, so that the split basis may be retained globally. 

Thus, we have translated all of the usual notions of gravitational theories (as presented in the first-order formalism) into the corresponding structures associated with a Lie algebroid. This again is an off-shell construction, and as such the off-shell symmetry has been imposed as a choice of $G$-structure, independent of the properties of any particular Lagrangian.

\section{Conclusions}\label{sec:concl}

Our aim in this work has been to set the right mathematical stage to formulate the geometric structure of gauge theories, including their BRST data. We have shown that this is the Atiyah Lie algebroid. This efficiently encodes the proper physical degrees of freedom, not only a gauge field but also a corresponding gauge ghost field.
Furthermore, we have understood carefully the exterior derivative on Lie algebroids and on arbitrary associated bundles, and demonstrated that it encodes two fundamental notions for gauge theories: the gauge covariant derivative in its horizontal part and the BRST operator $s$ in its vertical part. The understanding that BRST is imprinted in (Atiyah) Lie algebroids is novel, so we have taken time to show that it indeed gives rise locally to the well-known BRST transformations of the gauge field, the gauge ghost and charged matter fields.
Contrary to previous results in the literature, there is no need to impose external constraints to achieve these transformations, but instead they are just part of the geometry. 

It is important to realize that much of the mathematics and physics of the usual description of gauge theories involves not just a principal bundle but the tangent and cotangent bundles thereof.
A substantial difference from previous works on geometrization of gauge theories is that the Atiyah Lie algebroid is in fact a bundle over the space-time $M$ rather than the principal bundle $P$. 
We have already seen some payback of this fact: physical data like the gauge and ghost fields are encoded in bundle maps over $M$ itself, removing redundant group action.
We have concluded the body of the manuscript exhibiting how gravitational theories are treated on an equal footing to Yang-Mills theories in our formalism. The only extra ingredient is the solder form, which we have properly interpreted as related to a bundle associated to the Atiyah Lie algebroid. 

There are many possible applications and extensions of our work. First, since the construction is geometric, it is natural to apply the formalism to physical models, involving the specification of an action and quantization. This will presumably require extra structure, such as antighosts, which would be defined as additional fields on our geometric framework. Second, it is of interest to exploit the differential complex, for example in the context of the geometry of conformal anomalies, within the Lie algebroid context. 
In this manuscript we have confined our attention to transitive Lie algebroids. Extend our results to intransitive Lie algebroids, on top of being an interesting mathematical question, might be of  physical interest, as this would make contact with developments on Courant algebroids and Lie bialgebroids in the context of double field theory, metastring theory and generalized $T$-duality, see e.g. \cite{Gualtieri:2003dx, Grana:2008yw, Freidel:2015pka, Freidel:2017wst, Coimbra:2011nw, Vaisman:2012ke, Svoboda:2018rci} and references therein.

Another extension of our work that we expect to have many physical applications would involve Lie algebroids with boundaries, interfaces and corners on the base manifold $M$.
Indeed, this was one of our motivations for initiating the present work, as such constructions are clearly better formulated when all of the physics is associated with bundles over $M$.  Possible applications include holography, quantum symmetries of gravity, entanglement and asymptotic symmetries, which are all currently of intense interest. For the latter, it would be interesting to explore the consequences of our formalism to the variational bicomplex of Anderson \cite{MR1188434}. In this context, we regard the appearance of BRST symmetry built-in to the structure of Atiyah Lie algebroids as an indication that Lie algebroids should be used as the building blocks of a variational calculus for gauge theories. Indeed in either the classical or quantum domain, one expects to need to explore the space of connections (and other fields) on a Lie algebroid. In the present paper, we have simply worked at a single point in such a space, regarding the Ehresmann connection  as fixed but arbitrary.
We expect to explore this further in subsequent work.

\paragraph{Acknowledgements}

We are grateful to Glenn Barnich, Laurent Baulieu and Weizhen Jia for enlightening discussions.
LC is supported by the ERC Advanced Grant \textsl{High-Spin-Grav}. The work of RGL was supported by the U.S. Department of Energy under contract DE-SC0015655.
\appendix

\renewcommand{\theequation}{\thesection.\arabic{equation}}
\setcounter{equation}{0}

\section{Conventions}\label{sec:notation}

This Appendix is devoted to a summary of the  conventions used throughout the manuscript.

Given a bundle $B$ over a manifold $M$, the space of its local sections is denoted $\Gamma(B)$. Sections of a vector bundle are written with an underline, e.g., $\un\mX,\un\mY,...$ for the bundle $A$ over $M$. Examples of bundles over $M$ in the paper are $A$, $L$, $E$ and of course $TM$. We use the underline to distinguish these sections from functions in $C^\infty (M)$ or the components of the sections, which are a set of $rank\ B$ functions. One of the advantages of this notation is that, as long as there are no explicit indices in an equation the latter is completely invariant (tensorial). Locally, we can introduce a basis of sections of a bundle. We carry over the underline convention to the basis, to again distinguish it from the components. Consider for instance a vector bundle $L$ over $M$. Its sections are denoted $\un\mu$. We then introduce a basis $\lbrace \un t_A \rbrace$ with the index $A$ running over $rank\ L$ values and write $\un\mu=\mu^A \un t_A$. Here, $\mu^A$ are just $rank\ L$ functions of $M$. The utility of this convention is clear once we consider maps between bundles over $M$ with pre-image $L$. An example is the map $\iota$ mapping $L$ to $A$, which is another bundle over $M$. While in index-free notation $\iota(\un\mu)$ is just a section of $A$, we can write 
\beq
\iota(\un \mu)=\iota(\mu^A\un t_A)=\mu^A \iota(\un t_A),
\eeq 
where we simply used the fact that $\iota$ is linear with respect to multiplication by  $C^\infty (M)$ functions. Having underlined the basis facilitates the employment of linearity of bundle maps to reduce them to maps between bases. Indeed, given a basis $\lbrace \un E_M \rbrace$ for sections of $A$, with the index $M$ spanning $rank\ A$ values, we have $\iota(\un t_A)=\iota^{M}{}_A \un E_M$, where $\iota^M{}_A$ are the components of the map $\iota$ in the specific bases for $L$ and $A$. For bundle maps, we conventionally write the target space index first and the pre-image space index second, as we did for $\iota^M{}_A$. All in all we thus have
\beq
\iota(\un \mu)=\iota(\mu^A\un t_A)=\mu^A \iota(\un t_A)=\iota^{M}{}_A\mu^A  \un E_M,
\eeq
where again the underline convention facilitates recognizing that the output is a section of $A$ expressed in the local basis $\lbrace \un E_M\rbrace$. 

Given a vector bundle $B$ over $M$, we have a dual bundle $B^*$, whose sections are  maps from $B$ to $C^\infty(M)$. To avoid confusion, we do not underline bases of $B^*$. Again, the most familiar example is $T^*M$; here we will use as specific example the vector bundle $E$. Once a basis $\lbrace \un e_a\rbrace$ ($a=1,..,rank\ E$) is specified, the dual basis of $E^*$ is then denoted $\lbrace f^b\rbrace$, defined via $f^b(\un e_a)=\delta^b{}_a$. Although there could be potential tension between the basis of $E^*$, $f^a$, and the components $\psi^a$ of a section of $E$, $\un\psi=\psi^a\un e_a$, we decide not to underline $f^a$ but rather to simply use different letters from the ones chosen for the components $\psi^a$. We refer to sections of $A^*$ as extended forms, and introduce the basis $\lbrace E^M \rbrace$ via $E^M(\un E_N)=\delta^M{}_N$. Extended forms in $A^*$ may be valued in associated bundles. Therefore, $\secE_n \in \Gamma(\wedge^n A^*\times E)$ is an $E$-valued extended $n$-form, mapping $n$ sections of $A$ to $E$. The underline in 
$\secE_n$ is due to the fact that $\secE_n(\un \mX^1,...,\un \mX^n)$, with $\un\mX^1,...,\un\mX^n \in \Gamma(A)$, is a section of $E$. Thus, we write expressions like $\secE_1(\un \mY)$, which simply means that the $E$-valued extended one form $\secE_1$ acts on a section of $A$ to produce a section of $E$. 
Note that since $\hatd$ is defined to act on sections of $\wedge^nA^*\times E$ via the morphism $\phi_E$, some authors include a subscript on $\hatd$ to emphasize this. Here we have chosen not to, so that $\hatd$ should be understood by context; see for example eq. \eqref{dpsi}. In this equation we also introduce the notation $\nabla^{\wedge^nA^*\times E}$; here we denote by $\nabla^B$ the induced connection for any bundle $B$ associated to $A$. As this might be difficult to understand at first, we expand on its meaning in App. \ref{app:LAmaps}. Extended forms on $A$ can be seen equivalently as quantities living in $\wedge^nA^*\times E$, or as maps taking $n$ sections of $A$ to $E$. It is important therefore to extract from the resulting section of $E$ the form part acting on sections of $A$. As an illustrative example, consider $\hat d \un\psi_1 \in \Gamma(\wedge^2 A^*\times E)$. Acting on two sections $\un\mX$ and $\un\mY$ of $A$ it produces a section of $E$, $
\hat d \un\psi_1 (\un\mX,\un\mY).$ 
Moreover, one can extract from this the extended two form 
\beq\label{extract}
\hat d\un\psi_1=(\hat d\psi_1)^a_{MN} E^M\wedge E^N \otimes \un e_a,
\eeq
where $(\hat d\psi_1)^a_{MN}$ are just the components of this object living in $\Gamma(\wedge^2 A^*\times E)$. This process is crucial in understanding the covariant derivative and BRST action. To characterize them, we furthermore need to discuss how to extract the horizontal and vertical parts of the extended forms themselves. This is established in Sec. \ref{sec:derivations} and further commented upon in App. \ref{app:LAmaps}.

\section{$\hatd$ and extended forms}\label{app:LAmaps}
\setcounter{equation}{0}

We have computed various aspects of the $\hat d$-action. Since formulae can be rather involved, in this Appendix we give full details of how $\hatd$ acts  on extended one forms valued in $E$, i.e., on sections $\Gamma(A^*\times E)$. We have written the result in a compact form in the text by using the induced connection on $A^*\times E$. Recall that given a connection $\nabla^B$ on a vector bundle $B$, there is a connection induced on $B^*$ via
\beq
\un X(\alpha(\un v))=(\nabla^{B^*}_{\un X}\alpha)(\un v)+\alpha(\nabla^{B}_{\un X}\un v),\qquad \un X\in\Gamma(TM), \alpha\in\Gamma(B^*), \un v\in\Gamma(B),
\eeq
which follows from Leibniz. 
For example, given \eqref{conncoeffsE}, we have immediately
\beq
\nabla^{E^*}_{\un\mX_H}f^a=-A^a{}_b(\un\mX_H)f^b,
\eeq
since $f^a(\un e_b)=\delta^a{}_b$ is constant. The connections on $\wedge^n A^*\times E$ are deduced similarly, based on Leibniz. Here, we would like to give some details on extracting the horizontal-vertical split of $\hatd\secE_1$ as an example,
\beq\label{n1}
\hatd\secE_1=\nabla^{A^*\times E}\secE_1+s\secE_1,
\eeq
from the Koszul formula \eqref{koszulE}. Note that this process is not as immediate as it was for sections of $E$ since $\hatd\secE_1(\un\mX,\un\mY)$ involves {\it two} sections each of which have horizontal and vertical parts. The Koszul formula gives
\beqn\label{dpsi1xy}
\hat d\un\psi_1 (\un\mX,\un \mY)
&=&
\phi_E(\un\mX)(\secE_1(\un\mY))-\phi_E(\un\mY)(\secE_1(\un\mX))
-\secE_1(\big[\un\mX,\un\mY\big]_A)
\nonumber
\\
&=&
(\hatd(\secE_1(\un\mY)))(\un\mX)-(\hatd(\secE_1(\un\mX)))(\un\mY)
-\secE_1(\big[\un\mX,\un\mY\big]_A)
\nonumber
\\
&=&
\nabla^E_{\un\mX_H}(\secE_1(\un\mY))
-\nabla^E_{\un\mY_H}(\secE_1(\un\mX))
-\secE_1(\big[\un\mX,\un\mY\big]_A)
\nonumber
\\&&
-v_E\circ\omega(\un\mX_V)(\secE_1(\un\mY))
+v_E\circ\omega(\un\mY_V)(\secE_1(\un\mX)),
\eeqn
which follows since $\secE_1(\un\mX)\in\Gamma(E)$. We now split the sections $\un\mX,\un \mY$ into horizontal and vertical parts,
\beqn
\hat d\un\psi_1 (\un\mX,\un \mY)
&=&
\nabla^E_{\un\mX_H}(\secE_1(\un\mY_H))
-\nabla^E_{\un\mY_H}(\secE_1(\un\mX_H))
-\secE_1(\big[\un\mX_H,\un\mY_H\big]_A)
\nonumber
\\&&
+\nabla^E_{\un\mX_H}(\secE_1(\un\mY_V))
-\nabla^E_{\un\mY_H}(\secE_1(\un\mX_V))
-\secE_1(\big[\un\mX_H,\un\mY_V\big]_A)
+\secE_1(\big[\un\mY_H,\un\mX_V\big]_A)
\nonumber
\\&&
-v_E\circ\omega(\un\mX_V)(\secE_1(\un\mY))
+v_E\circ\omega(\un\mY_V)(\secE_1(\un\mX))
-\secE_1(\big[\un\mX_V,\un\mY_V\big]_A).
\eeqn
Here, we organized the right-hand side in three separately tensorial expressions.
Then, defining
\beq\label{nabAE0}
(\nabla^{A^*\times E}\secE_1)(\un\mX_H,\un\mY_H)
&=&\nabla^E_{\un\mX_H}(\secE_1(\un\mY_H))
-\nabla^E_{\un\mY_H}(\secE_1(\un\mX_H))
-\secE_1(\big[\un\mX_H,\un\mY_H\big]_A)\\
(\nabla^{A^*\times E}\secE_1)(\un\mX_H,\un\mY_V)&=&
\nabla^E_{\un\mX_H}(\secE_1(\un\mY_V))
-\secE_1(\big[\un\mX_H,\un\mY_V\big]_A)\\
(\nabla^{A^*\times E}\secE_1)(\un\mX_V,\un\mY_H) 
&=&-\nabla^E_{\un\mY_H}(\secE_1(\un\mX_V))
+\secE_1(\big[\un\mY_H,\un\mX_V\big]_A)\\
(\nabla^{A^*\times E}\secE_1)(\un\mX_V,\un\mY_V) 
&=&\un 0,\label{nabAE}
\eeqn
we have
\beqn
\hat d\un\psi_1 (\un\mX,\un \mY)
=
(\nabla^{A^*\times E}\secE_1)(\un\mX,\un\mY)
-v_E\circ\omega(\un\mX_V)(\secE_1(\un\mY))
+v_E\circ\omega(\un\mY_V)(\secE_1(\un\mX))
-\secE_1(\big[\un\mX_V,\un\mY_V\big]_A).
\eeqn
Comparing with \eqref{n1}, we then identify
\beqn
s\secE_1(\un\mX,\un\mY)&\equiv&-v_E\circ\omega(\un\mX_V)(\secE_1(\un\mY))
+v_E\circ\omega(\un\mY_V)(\secE_1(\un\mX))
-\secE_1(\big[\un\mX_V,\un\mY_V\big]_A)\\
&=& 
-\ghost_E(\secE_1(\un\mY))(\un\mX_V)
+\ghost_E(\secE_1(\un\mX))(\un\mY_V)
-\secE_1(\big[\un\mX_V,\un\mY_V\big]_A),
\eeqn
where we introduced $c_E$ using \eqref{brstsecE}.

Notice finally that (\ref{nabAE0}-\ref{nabAE}) give us an explicit formula for \eqref{gencovder}:
\beqn
(\nabla^{A^*\times E}\secE_1)(\un \mX,\un\mY)&=&(\nabla^{A^*\times E}_{\un\mX_H}\secE_1)(\un \mY)-(\nabla^{A^*\times E}_{\un\mY_H}\secE_1)(\un \mX)\label{nabAETot}\\
&=&\nabla^E_{\un\mX_H}(\secE_1(\un\mY))
-\nabla^E_{\un\mY_H}(\secE_1(\un\mX))
-\secE_1(\big[\un\mX_H,\un\mY_H\big]_A)\nonumber
-\secE_1(\big[\un\mX_H,\un\mY_V\big]_A)
+\secE_1(\big[\un\mY_H,\un\mX_V\big]_A).
\eeqn
This result is instrumental in the derivation of the BRST transformation of the gauge field, eq. \eqref{BRSTA}.

\providecommand{\href}[2]{#2}\begingroup\raggedright\endgroup
\end{document}